# Dynamic Switching Models for Truck-only Delivery and Drone-assisted Truck Delivery under Demand Uncertainty

Jiaqing Lu
Department of Civil and Environmental Engineering,
Florida State University
2525 Pottsdamer St, Tallahassee, FL 32310
jl23br@fsu.edu

Qianwen (Vivian) Guo, Ph.D. (corresponding author)
Department of Civil and Environmental Engineering,
Florida State University
2525 Pottsdamer St, Tallahassee, FL 32310
qguo@fsu.edu

Dian Sheng
School of Management
Huazhong University of Science and Technology
Wuhan 430074, China
diansheng@hust.edu.cn

Shumin Chen, Ph.D.
School of Management,
Guangdong University of Technology
No. 161, Yinglong Road, Guangzhou, 510520, China
chenshumin1@gmail.com

Paul Schonfeld, Ph.D.
Department of Civil and Environmental Engineering,
University of Maryland, College Park
1173 Glenn Martin Hall, College Park, MD 20742 US
pschon@umd.edu

# Abstract

Integrating drones into truck delivery systems holds the potential for transformative improvements in customer accessibility, operational cost reduction, and delivery efficiency. However, this integration comes with costs, including drone procurement, maintenance, and energy consumption. The decision on whether and when to incorporate drones into truck delivery systems is heavily contingent on the demand for these services. In areas where demand is low and dispersed, deploying drone-assisted trucks may lead to the underutilization of resources and financial challenges, primarily due to the substantial upfront costs of drone deployment. Accurately predicting future demand density is a complex task, compounded by uncertainties stemming from unforeseen events or infrastructure disruptions. To tackle this challenge, a market entry and exit real option approach has been used to optimize the switching timing between different delivery services (i.e., trucks with or without drones) while considering the stochastic nature of demand density. The results of this study highlight that drone-assisted truck delivery, particularly when multiple drones are deployed per truck, can offer significant economic advantages in regions with high demand for delivery services. Utilizing the proposed dynamic switching model, the deterministic and stochastic approaches reduce costs by 17.4% and 31.3%, respectively, compared to the immediate cost-saving switching model. Furthermore, the stochastic parameters within the real option framework asymmetrically influence the entry and exit timings, as revealed through sensitivity analysis. Then, a stochastic multiple-options model is developed to dynamically switch between truck-only delivery and drone-assisted truck delivery with varying numbers of drones. These proposed stochastic models are applied in the Miami-Dade County region to evaluate the cost of dynamic switching services for three major logistics operators in a real-world scenario. This study illuminates the potential benefits of dynamic switching between different delivery services and provides decision-makers in the logistics industry with valuable insights for optimizing their delivery systems.

# 1. Introduction

The incorporation of unmanned aerial vehicles (UAVs), or drones, into conventional truck-based delivery systems represents a significant paradigm shift for the logistics industry (Straubinger et al., 2021). This innovative approach has already been adopted by several major logistics companies such as UPS, Amazon, Walmart, and DHL (Macrina et al., 2020). The deployment of drones in last-mile delivery has the potential to reduce operational cost by offering more direct and energy-efficient travel, thereby mitigating environmental impacts compared to traditional ground-based delivery methods (Figliozzi, 2017; Dukkanci et al., 2021). Moreover, drone delivery extends service reach to remote or otherwise inaccessible regions, addressing logistical barrier in rural areas and during disaster response scenarios (Chowdhury et al., 2021; Ghelichi et al., 2022).

Despite these advantages, the widespread adoption of drone delivery faces several operational constraints. Drones typically accommodate only one package at a time due to payload limitations, and their flight range is restricted by battery capacity (Choi and Schonfeld, 2018; Liu et al., 2023). In contrast, conventional trucks offer large payload capacities and long-distance service capabilities, maintaining their dominance in delivery operations (Li et al., 2022). However, truck-only delivery systems incur higher labor costs and longer delivery times. To address these challenges, the drone-assisted truck delivery model has emerged, combining the complementary strengths of both modes while minimizing their respective limitations (Rejeb et al., 2023).

This hybrid model functions similarly to a feeder-truck system, where a delivery truck carries multiple drones that are deployed en route to serve nearby customers. Drones are launched from the truck to deliver packages and then return to a designated rendezvous point. Meanwhile, the truck continues along its route, potentially reuniting with the drones for recharging or redeployment (Yin et al., 2023). The feasibility of this model has been demonstrated by logistics providers. For instance, UPS has reported that drone-assisted truck systems, as depicted in Figure 1, potentially save up to $50 million annually by reducing just one mile from each of its 66,000 daily driver routes (Stewart, 2017). Similarly, Workhorse's Horsefly drone, mounted on its C-1000 electric truck, offers delivery costs as low as 3 cents per mile, which is an 80% reduction compared to traditional delivery expenses (Adler, 2020).

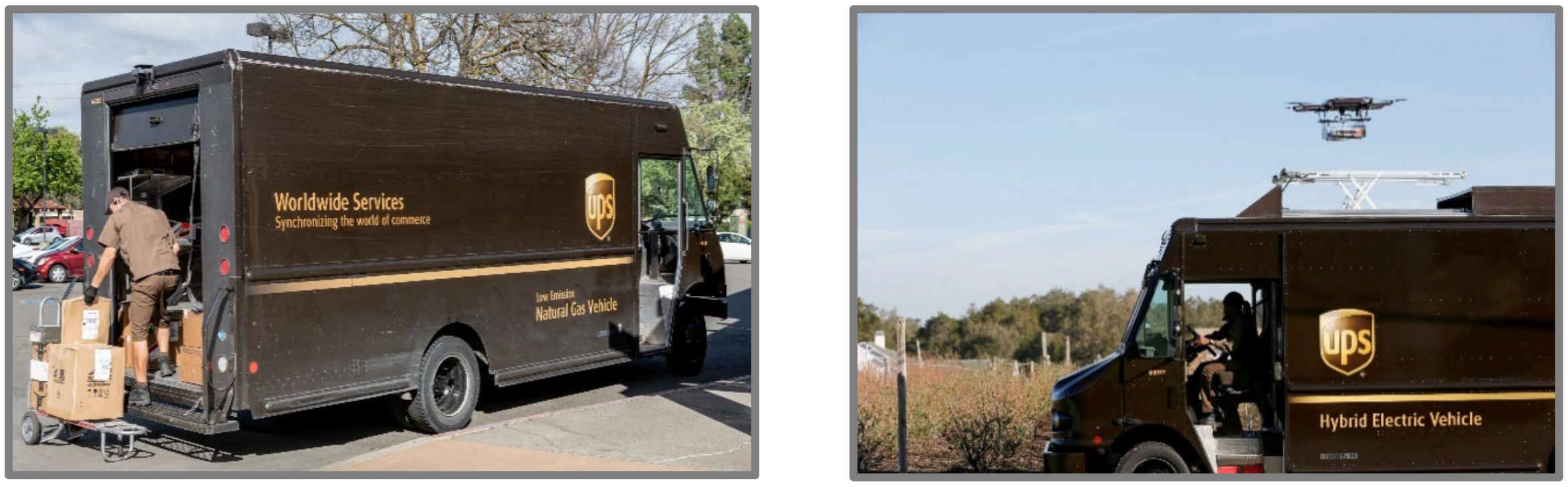

a) Truck-only delivery b) Drone-assisted truck delivery

**Figure 1.** Application of two delivery services (www.ups.com).

Nowadays, major carriers such as USPS, UPS, and FedEx maintain mixed fleets comprising both conventional trucks and drone-assisted trucks. A key operational challenge lies in determining how to dynamically allocate these resources in response to fluctuating demand densities (Kitjacharoenchai et al., 2019; Salama and Srinivas, 2022; Jeong and Lee, 2023). This study addresses the challenge by focusing on optimizing switching strategy between truck-only and drone-assisted truck delivery modes to minimize operational costs. As illustrated in Figure 2, the two delivery modes differ significantly: in the drone-assisted mode, trucks and drones operate in parallel to serve multiple delivery points simultaneously, while in the truck-only mode, all deliveries are completed sequentially by the truck alone.

While many logistics companies employ both delivery modes, the allocation of resources often overlooks fluctuating demand levels. Each mode performs optimally under different demand conditions, and relying solely on one may lead to inefficiencies such as underutilized capacity or increased costs. To this end, the study first develops detailed cost functions for each delivery mode to determine the demand thresholds at which switching becomes advantageous. Since delivery demand is inherently uncertain, being affected by seasonal fluctuations, urban congestion, and consumer behavior, rigid fleet allocation strategies may result in inefficiencies. A flexible, demand-responsive approach is thus desirable for enabling timely switching between delivery modes and improving overall system efficiency.

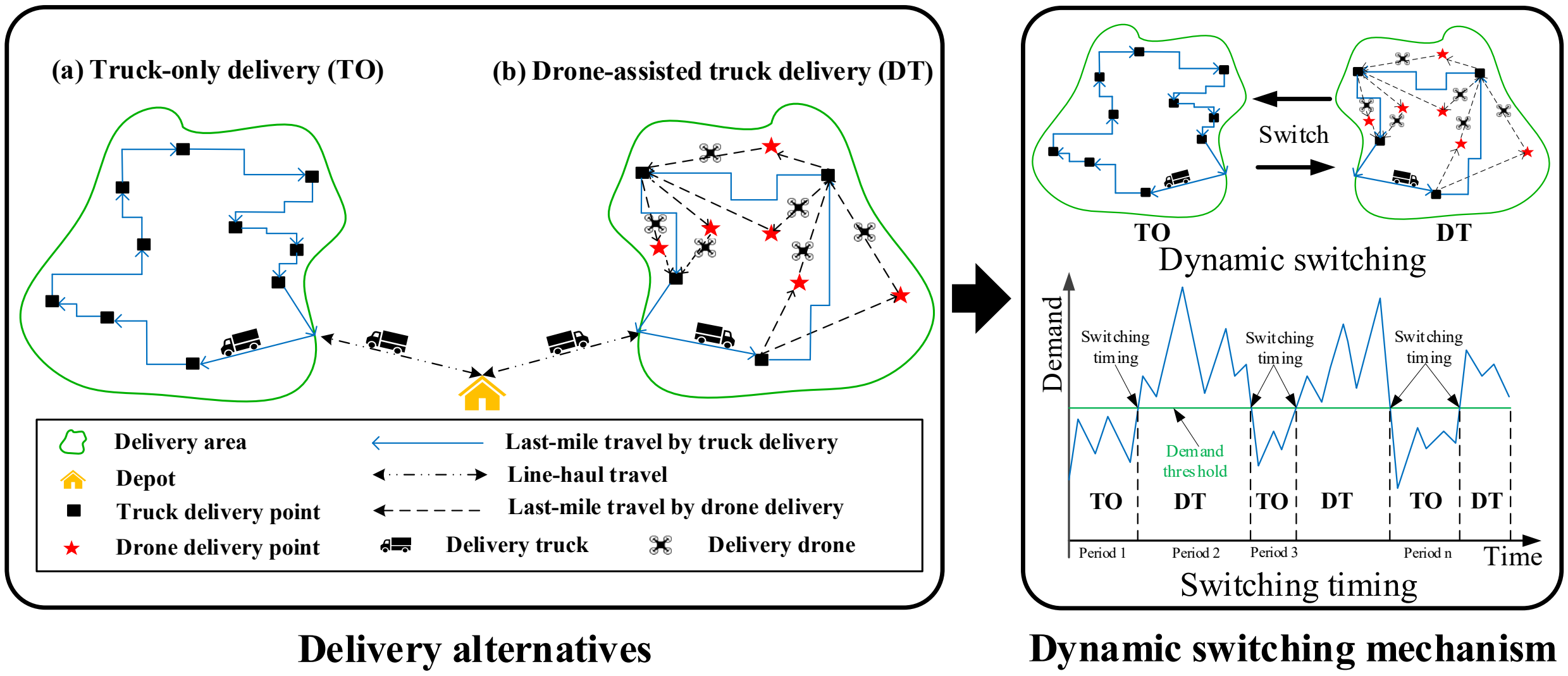

**Figure 2.** Overview of dynamic delivery mode switching between TO and DT systems.

Existing literature has largely focused on optimizing delivery routes and operations under static demand assumptions (Murray and Chu, 2015; Ha et al., 2018; Karak and Abdelghany, 2019; Schermer et al., 2019). In contrast, this study introduces a dynamic mode-switching model that explicitly accounts for demand uncertainty. To capture this uncertainty, the research adopts an entry-exit real options framework, where future demand density follows a Geometric Brownian Motion (GBM) process. Originally introduced to model market entry and exit decisions under price uncertainty (Dixit, 1989; Dixit and Pindyck, 1994), this framework has since been applied in various domains, including freight shipping (Balliauw, 2017; Sødal et al., 2008; Sødal et al., 2014), autonomous vehicle operations (Guo et al., 2017), mining strategies (Marmer and Slade, 2018), and investments in energy conservation (Lin and Huang, 2010). The entry-exit real option approach is well-suited for the last-mile delivery context, offering logistics operator the flexibility to switch between services based on stochastic demand conditions. Here, integrating drones into delivery operations is modeled as an "entry" decision, while reverting to truck-only operations represents an "exit." This modeling approach enables logistics decision-makers to assess the value and optimal timing of switching delivery modes in a manner that reflects both economic trade-offs and demand uncertainty. The resulting framework provides a robust foundation for enhancing delivery system performance in dynamic, uncertain environments.

The main technical aspects this study are four-fold. First, detailed operational cost functions are formulated for both truck-only and drone-assisted truck delivery modes, enabling a quantitative evaluation of the economic trade-offs involved in switching between these alternatives. Second, a stochastic dynamic switching model is developed using an entry-exit real options framework to identify the optimal timing for switching between conventional and drone assisted truck deliveries. In contrast to traditional models that assume the steady demand, this approach explicitly incorporates demand uncertainty. Third, the framework is extended to a multi-option switching model to accommodate a wider range of operational scenarios. Finally, the proposed models are validated through a case study conducted for the Miami-Dade County (MDC) area. This application evaluates the effectiveness of dynamic switching strategies for three major logistics operators, demonstrating the practical utility of the models and revealing potential cost savings compared to traditional mode operations.

The subsequent sections of this article are organized as follows. Section 2 presents the literature review on delivery service evolution and real options applications in transportation. In Section 3, cost functions for two distinct types of delivery services are specified. Section 4 develops the switching timing models, encompassing deterministic and stochastic demand settings. Section 5 presents the findings from the immediate cost-saving (IC) switching model, the deterministic model, and the stochastic model, along with assessments of how these results are affected by various significant parameters. The proposed model is applied in the context of Miami-Dade County, where we conduct simulations to determine the optimal time periods for two types of delivery services for three logistics operators. Lastly, Section 6 provides the conclusions drawn from this study and outlines potential research directions for the future.

# 2. Literature Review

Drone-assisted truck delivery has attracted significant interest from both industry and academia due to its potential to revolutionize last-mile logistics. Prior studies have highlighted its operational benefits, including reduced delivery times (Wang et al., 2016) and lower overall costs and emissions compared to traditional truck-only methods (Meng et al., 2023). To address the routing complexities of these hybrid systems, researchers have developed specialized models such as the Flying Sidekick Traveling Salesman Problem (Murray and Chu, 2015) and employed Continuous Approximation (CA) techniques for optimization (Campbell et al., 2017; Carlsson and Song, 2018). Despite these advancements, real-world deployment remains limited, primarily due to demand uncertainty, which undermines the performance of static deployment strategies. In response, the real options (RO) approach has emerged as a promising framework for supporting decisions under uncertainty. Zheng and Jiang (2023) conducted a comprehensive review that underscores the growing application of RO in dynamic transportation settings. Nevertheless, a critical gap remains in identifying optimal conditions and timing for transitioning between truck-only and drone-assisted modes. This paper's literature review focuses on research relevant to the dynamic mode-switching problem in drone-assisted truck delivery systems under uncertainty. The reviewed studies are categorized into three main areas: (1) comparisons of last-mile delivery strategies and optimization of drone-truck operations, (2) applications of switching models in various transportation contexts, and (3) utilization of real options approaches to manage demand uncertainty in strategic decision-making. Together, these strands of literature provide the foundation for our proposed dynamic switching framework.

## 2.1 Drone-assisted truck delivery models

Compared to traditional truck-only delivery services, integrating drones into existing delivery systems offers transformative potential in terms of customer accessibility, cost-efficiency, and overall delivery performance (She and Ouyang, 2021; She and Ouyang, 2024). While drone-assisted delivery incurs additional costs related to drone acquisition and maintenance (Chen et al., 2021), it can significantly reduce delivery times by deploying drones directly from trucks (Mourelo Ferrandez et al., 2016). In scenarios with high-speed or multiple drones, Wang et al. (2016) projected a significant reduction of 75% in roundtrip times compared to unassisted trucks. Campbell et al. (2017) recognized drone-assisted truck deliveries as ideal for dense demand areas, given the balance of their operational and stoppage costs. Meng et al. (2023) further demonstrated environmental and economic benefits, reporting a 24.90% reduction in carbon emissions, a 22.13% decrease in total costs, and a 20.65% reduction in delivery times

However, most of the above studies focus on static settings, overlooking the dynamic challenges faced in real-world logistics, such as uncertainty stemming from weather, traffic, and fluctuating customer demand. This paper focuses on uncertainty in future demand density, a key factor in determining whether and when to deploy drone-assisted truck services. In parallel with the exploration of drone-assisted benefits, substantial research has addressed the routing complexities introduced by these hybrid systems. Routing problems traditionally rely on formulations like the Traveling Salesman Problem (TSP) and the Vehicle Routing Problem (VRP) (Fan et al., 2024; Ouyang, 2007; Li and Quadrifoglio et al., 2011; Zhen and Gu, 2024; Shahin et al., 2024), which aim to minimize customer wait times and operational costs by optimizing delivery sequences (Robusté et al., 2004; Figliozzi, 2009). These models have evolved into more sophisticated variants, such as the Cumulative Capacitated VRP (Ngueveu et al., 2010) and the TSP with profit or time-window constraints (Bjelić et al., 2013; Dewilde et al., 2013).

The integration of drones into delivery workflows has driven further development of specialized routing models. Murray and Chu (2015) introduced the Flying Sidekick Traveling Salesman Problem (FSTSP), a pioneering model where a drone is dispatched from a truck to deliver to a customer while the truck continues en route. This model aims to reduce joint truck-drone delivery times. Other adaptations include the TSP with Drones (TSP-D), the Multiple TSP with Drones, and the Vehicle Routing Problem with Drones (VRP-D), all aiming to optimize cost and service time (Lei and Chen, 2022). Sacramento et al. (2019) proposed an adaptive large neighborhood search algorithm for the VRP-D, while Tiniç et al. (2023) extended the TSP-D to scenarios involving multiple drones per truck.

In contrast to discrete optimization models, Daganzo (1984a) pioneered the Continuous Approximation (CA) method to estimate routing and travel distances in large-scale transportation systems. This model has been extensively used in logistics optimization, including depot placement (Ouyang and Daganzo, 2006; Stokkink and Geroliminis, 2021) and delivery network design (Ansari et al., 2018; Zhang et al., 2024; Stokkink et al., 2024). Campbell et al. (2017) applied Daganzo's swath model within a CA framework to address hybrid delivery routing challenges, and Carlsson and Song (2018) extended this approach to minimize last-mile delivery time in truck-drone systems. Chowdhury et al. (2021) used CA to evaluate drone effectiveness in post-disaster delivery scenarios where truck routes were blocked.

While an extensive literature compares static drone-assisted and conventional delivery models, limited attention has been paid to the dynamic decision of when to switch between these modes in response to uncertain demand. This study aims to fill this gap by formulating cost functions for both delivery modes and proposing a dynamic switching model to support operational decisions under uncertainty.

## 2.2 Switching models in the transportation field

Switching models have been widely applied in transportation, with early applications in maritime shipping and more recent extensions to multimodal and fleet management contexts. In maritime shipping, research has primarily addressed optimal timing for carriers to switch between dry bulk and wet bulk markets, leveraging market dynamics and real options to enhance operational flexibility and profitability. Beenstock and Vergottis (1993) introduced a pioneering model in which combined carriers transition between tanker and dry cargo trades based on profitability, highlighting substantial spillover and feedback effects across shipbuilding, freight, and scrap markets. Brekke and Øksendal (1994) extended the theoretical framework by formulating a switching model based on impulse control and quasi-variational inequalities to optimize the timing of entry and exit in resource extraction under stochastic price dynamics. Sødal et al. (2008) built on these insights with an entry-exit switching model that quantifies the strategic value of switching for shipping firms, and their subsequent work (2014) demonstrated that although the second-hand ship market generally adjusts efficiently, delayed market expectations around the early 2000s created notable profit opportunities. Adland et al. (2017) further extended the analysis to geographic switching options for dry bulk vessels and separately examined the seaborne oil transportation sector. Their study demonstrated that the value of switching between crude and product tankers can surpass the additional cost of constructing a specialized product tanker.

Beyond maritime applications, switching models have been increasingly applied to transit and multimodal transport systems to optimize service delivery under uncertainty. In fleet management, Guo et al. (2017) developed a real options approach for dynamically transitioning between fixed-route and flexible transit services, demonstrating substantial cost savings over static policies. Kim et al. (2018) developed a switching strategy between fixed-route and flexible-route transit services in a many-to-one demand setting, employing analytical optimization and a genetic algorithm for problem-solving. In multimodal transport, Lemmens et al. (2019) introduced a synchromodal decision rule that enables real-time shifts among coexisting transport modes to boost efficiency and sustainability. Sayarshad and Gao (2020) applied a queuing-based switching model to optimize the handoff between fixed-route and on-demand transit, improving both social welfare and last-mile performance. Most recently, Tang et al. (2023) designed a learning-enhanced framework that integrates flexible electric buses with fixed-route networks, jointly optimizing routing, scheduling, and transfers to enhance overall system efficiency.

Despite these advances, the rapid growth of drone-based logistics and e-commerce highlights an urgent need for real-time switching models tailored to express-delivery vehicle fleets, a gap this paper aims to address.

## 2.3 Real options approach applications for addressing demand uncertainty

Demand uncertainty poses significant challenges in investment and operational decision-making across various industries, including transportation, energy, and manufacturing (Zheng and Jiang, 2023). Traditional net present value methods often fail to capture the flexibility required to respond to uncertain market conditions. The real options approach is widely acknowledged as an effective strategy for dynamic switching problems when conducting a cost-benefit analysis prior to making an implementation decision. Real options have been elaborated upon by notable researchers (Trigeorgis, 1993; Dixit and Pindyck 1994; Trigeorgis and Reuer, 2017). This methodology has found fruitful application in diverse transportation field such as aviation investment (Xiao et al., 2017; Zheng et al., 2020; Guo and Jiang, 2022), maritime operation (Rau and Spinler, 2016; Zheng and Chen, 2018), and transit planning (Chow et al., 2011; Chow et al., 2016; Guo et al., 2017; Guo et al., 2023).

Within the real options framework, Dixit (1989) introduced an entry-exit model based on the Geometric Brownian Motion (GBM) process, highlighting the value of investment under demand uncertainty. GBM, a stochastic process, effectively models random variations in demand over time (Mahnovski, 2006). Brandão and Saraiva (2008) evaluated investments in PPP transit projects, emphasizing risk-sharing mechanisms and adaptive contracts. Fernandes et al. (2011) analyzed switching options in energy projects, accounting for fuel price volatility and technological advancements. Using GBM-driven simulation uncertainty, Zheng and Chen (2018) developed a real options model with a least squares Monte Carlo simulation to optimize fleet replacement timing under passenger demand and fuel price uncertainty. Zheng et al. (2020) applied a real options approach to airlines' investment timing in exclusive airport facilities under passenger demand ambiguity, identifying optimal investment rules and evaluating lump-sum and per-unit subsidies to align private and social investment decisions. Zheng et al. (2021) examined the timing decisions of two competing shipping lines for dedicated terminal investments under shipping demand ambiguity. Guo et al. (2021) presented a real options-based capacity integration model for port clusters and optimized multistage capacity investment and exit decisions under demand and congestion uncertainty, with subsequent enhancements improving social welfare and resource efficiency.

Despite this breadth, few studies address real-time switching between service modes in last-mile logistics. The entry-exit option perfectly suits scenarios where operators can activate or deactivate delivery modes, such as conventional trucks versus drone-assisted trucks, in response to evolving demand. To fill this gap, this research develop analytical optimization models for drone-assisted truck delivery under demand uncertainty, using a GBM process to capture nonstationary stochastic demand. The framework quantifies the value of flexibility: not only determining when to switch modes but also when to postpone that decision to maximize expected returns in an uncertain environment.

### 2.4 Summary

After reviewing an extensive body of research on delivery methods, it becomes evident that no studies have delved into the crucial aspect of switching timings between truck-only and drone-assisted-truck to optimize a delivery system. Moreover, most previous studies have focused on scenarios with static and deterministic demand density, with a notable absence of research considering demand uncertainty in the context of drone-assisted truck deliveries. To address these shortcomings, this study proposes a dynamic, stochastic switching model that determines when to deploy traditional trucks versus drone-assisted trucks under demand uncertainty. By modeling demand as a GBM, this approach captures real-world fluctuations in delivery density. The adoption of the entry-exit real option approach proves particularly fitting for the last-mile delivery service switching problem due to its capacity to navigate dynamic market conditions. This flexibility enables decision-makers to time mode switches strategically, adapting to shifting market conditions and maximizing expected returns in the face of uncertainty.

## 3. Operation Cost Formulation

This section formulates the operational costs for two delivery modes: truck-only (TO) and drone-assisted truck (DT), by quantifying their expected travel distances and associated cost components. It is assumed that customer points are independently and uniformly distributed at densityof $Q$ over a service zone $A$ with a centrally located depot (Figure 3). In both modes, the truck incurs a fixed "linehaul" cost to reach the delivery swath. Under TO, the truck alone performs every task, driving on rectilinear routes, parking, loading and unloading, and final drop-offs, incurring travel costs proportional to the expected rectilinear distance per point plus a fixed stop cost at each customer. Under DT, the truck travels the same linehaul distance and then either remains stationary or continues along its route while its onboard drones serve assigned customers via straight-line (Euclidean) flights. Customers are partitioned between the truck and the available drones to optimize last-mile efficiency, and a synchronization delay is included for the time the truck must wait for all drones to return before proceeding.

For each customer point, we break down the per-point cost into four components: (1) the truck's travel cost, allocated between the linehaul and local swath distances; (2) the drone travel cost based on the expected Euclidean flight distance; (3) fixed stop costs for truck visits and for each drone launch/recovery; and (4) the cost of synchronization time, valuing truck idle time at its operating rate. Summing these elements yields the expected cost per delivery point as a function of demand desity $Q$, the number of drones, vehicle speeds, unit travel and stop costs, and waiting-time valuation. Integrating this per-point cost over all delivery points produces the total operational cost for each mode. Finally, by comparing the TO and DT total costs, a cost-savings function is derived that identifies the demand density thresholds at which an instantaneous mode switch minimizes overall cost.

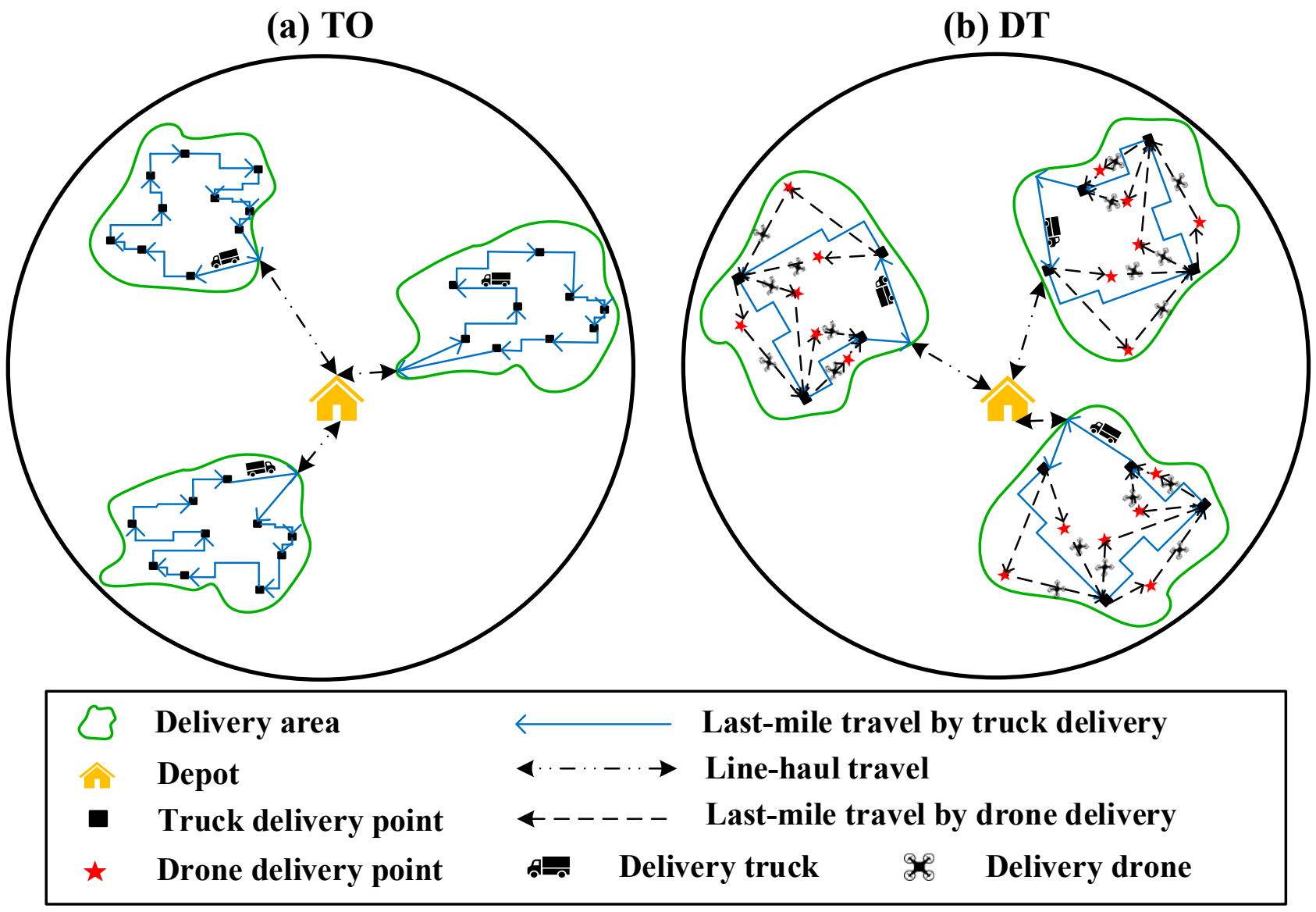

**Figure 3.** Service zone.

The notation and units for parameters used in the following analysis are shown in the following Table 1.

**Table 1.** Variable Definition.

| Parameters | Definitions |
|---|---|
| $A$ | Service zone |
| $C_t^o, C_d^o$ | Operation cost of truck/drone per unit distance |
| $C_{TO}, C_{DT}$ | Total cost of truck-only delivery in service zone |
| $d_t$ | Average stop time per truck delivery at a delivery point |
| $D_{TO}$ | Expected last-mile distance per delivery point of TO |
| $D_{DT}^t$ | Expected truck delivery distance per delivery point of DT |
| $D_{DT}^d$ | Total expected drone travel distance of DT |
| $E_{TO}, E_{DT}$ | Expected cost of TO/DT per delivery point |
| $H_T$ | Horizon planning time |
| $L$ | Expected straight linehaul distance |
| $m_{TO}, m_{DT}$ | The number of delivery points per route of TO/DT |
| $n$ | Number of drones per truck |
| $Q(0)$ | Demand density at time 0 |
| $S_t$ | Fixed truck stop cost |
| $S_d$ | Marginal cost for drone delivery, including launch and recovery cost, relative to truck delivery |
| $T$ | Expected route time |
| $V_l$ | Truck linehaul speed from depot to delivery area |
| $V_t, V_d$ | Truck/drone tour tour speed in delivery area |
| $w$ | Swath of width |
| $P_{TO}, P_{HD}, P_{DT}$ | Total cost of TO/HD/DT fleet to serve all delivery demand in whole region |
| $\varepsilon_{TO}, \varepsilon_{DT}^n$ | Proportion of customer points delivered by TO/ DT(n-drones) |
| $\phi, \nu$ | Circuity factor/adjust factor for line-haul distance |
| $\rho, \mu, \sigma$ | Annual discount rate, demand growth rate, and volatility rate |
| $F^+(F^-)$ | Switching cost |

## 3.1 Truck-only delivery

TO service operates as follows: the truck departs the central depot, follows a predetermined route through the service zone, visits each customer at least once to deliver an item, and then returns to the depot. As illustrated in Figure 3(a), the truck first travels from the depot into the service zone and proceeds to the designated delivery area. The final delivery leg carried out entirely by the driver encompasses loading and unloading goods, finding and maneuvering into a parking space, and completing the delivery at each customer location. In Figure 4(a), the truck's tour is depicted by the bold blue lines connecting successive delivery points.

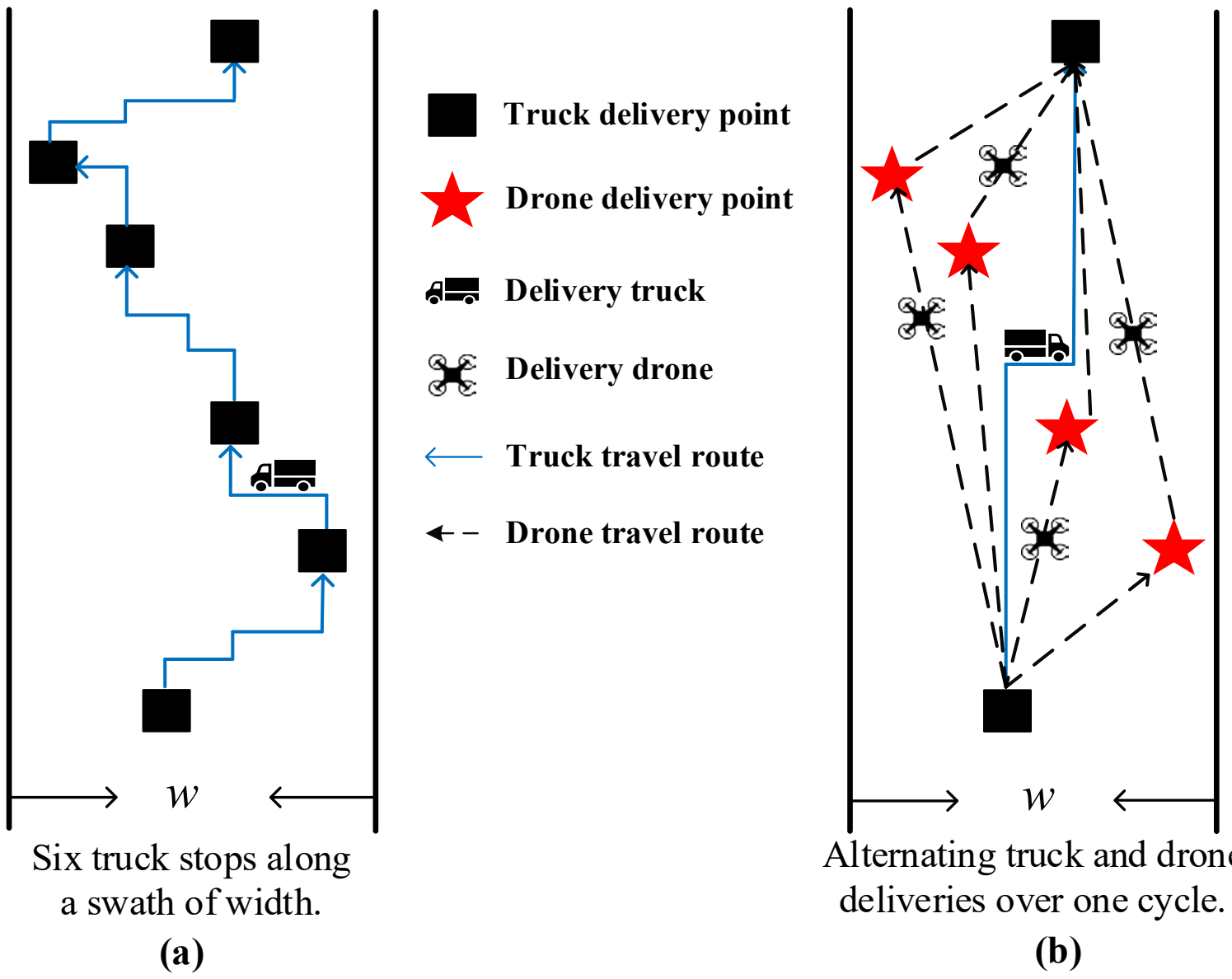

**Figure 4.** Truck and drone stop along a relatively narrow swath of width (delivery corridor/cycle).

Within the delivery area, we approximate the compact service region as a continuous-approximation swath of width $w$ (Daganzo, 1984b). In Figure 4(a), the truck travels back and forth along this swath, visiting customers in sequence according to their longitudinal positions.

The detailed derivation of the expected per-point cost for the truck-only (TO) mode is provided by Campbell et al. (2017) and summarized in Appendix G. Denoting the optimal per-customer cost by $E_{TO}$, it consistest of two components: the truck's variable cost and fixed stop cost at each delivery location:

$$E_{TO} = C_t^o * \left(D_{TO} + \frac{L}{m_{TO}}\right) + S_t, \quad (1)$$

where $C_t^o$ is the variable cost of truck per unit distance; $D_{TO}$ is the expected last-mile distance per delivery point; $m_{TO}$ is the number of delivery points per delivery route; $L$ is the linehaul distance and $S_t$ is the fixed truck stop cost per delivery point.

The total cost of TO delivery in the whole service zone $C_{TO}$ is then given by multiplying the per-point cost $E_{TO}$ by the total number of deliveries, $A * Q$:

$$C_{TO} = E_{TO} * A * Q = \alpha_1 Q + \beta_1 Q^{\frac{1}{2}}, \quad (2)$$

where $\alpha_1 = \left(\frac{L d_t C_t^o V_l}{V_l T - 0.9027\sqrt{A}} + s_t\right) A$, $\beta_1 = \left(\frac{2C_t^o}{\sqrt{3}} + \frac{2C_t^o V_l}{\sqrt{3} V_t \left(V_l T - 0.9027\sqrt{A}\right)}\right) A$.

## 3.2 Drone-assisted truck delivery

In DT delivery, a single truck and $n$ onboard drones collaborate to complete each "delivery cycle" along a swath of width $w$ (Figure 4(b)). The $n$ drones serve $n$ delivery points (depicted as red stars) positioned between two consecutive truck stops. The truck's route is shown with rectilinear (L1 metric) travel as bold blue lines, while the drones' paths are straight-line travel (L2 metric) indicated by black dashed lines. Before proceeding to the next $n+1^{st}$ delivery point (next cycle), the truck must wait for all drones to return, incurring a synchronization time at the end of each cycle. Therefore, the truck's estimated horizontal travel distance per point in one cycle is expected to be $\frac{w\left(1-\frac{1}{n+1}\right)}{3} = \frac{w}{3(n+1)}$, while the longitudinal travel distance remains consistent at $\frac{1}{Qw}$ due to the requirement of passing all stops in the longitudinal direction. Therefore, the expected truck delivery travel distance per delivery point $D_{DT}^{t}$ is:

$$D_{DT}^{t} = \frac{w}{3(n+1)} + \frac{1}{Qw}, \tag{3}$$

where the optimal swath width of expected truck delivery per delivery point $w^*$ is $\sqrt{\frac{3(n+1)}{Q}}$.

When $n$ drones are carried by one truck, the truck serves the $n+1^{st}$ delivery point and drones serve $n$ delivery points between two truck delivery points in one delivery cycle. The drones' fraction of served points is $\frac{n}{n+1}$. Then, the total expected drone travel distance per cycle is obtained by summing all drone delivery distances in this cycle $\sum_{i=1}^{n}\sqrt{\left(\frac{w}{3}\right)^2+\left(\frac{i}{Qw}\right)^2}+\sqrt{\left(\frac{w}{3}\right)^2+\left(\frac{n+1-i}{Qw}\right)^2}$.To simplify this equation, the distance for a drone delivery can be estimated as the average drone delivery distance. The longitudinal component of this total distance can be approximated with the half distance the between two truck stops, which is represented by $\frac{n+1}{2Qw}$. Therefore, the expected drone distance per delivery point $D_{DT}^{d}$ can be estimated in Eq. (10) through substituting the approximate longitudinal distance in the second term in each square root in Eq. (5) (Campbell et al., 2017).

$$D_{DT}^{d} = \frac{2n}{n+1}\sqrt{\left(\frac{w}{3}\right)^2+\left(\frac{n+1}{2Qw}\right)^2}, \tag{4}$$

The optimal swath width $w^*$ of Eq. (4) is approximated by $\sqrt{\frac{3(n+1)}{2Q}}$ by minizing expected drone travel distance, as detailed in Appendix H.

The expected linehaul travel distance per delivery point is $\frac{L}{m_{DT}}$, which depends on the number of delivery points for drone-assisted truck delivery $m_{DT}$ in Eq. (14). The operating cost of a drone encompasses the capital costs associated with infrastructure, drones, and batteries, as well as labor cost related to drone operations, maintenance, and electricity. Therefore, the expected cost of drone-assisted truck delivery per delivery point $E_{DT}$ includes truck travel cost, drone travel cost, stop cost for both truck and drones and synchronization time cost:

$$E_{DT} = C_t^o \left( D_{DT}^t + \frac{L}{m_{DT}} \right) + C_d^o D_{DT}^d + \frac{n}{n+1} S_d + S_t + C_s, \tag{5}$$

where $n$ is the number of drones per truck.

This waiting time, referred to as the synchronization time cost, depends on the distance traveled by the drones in a delivery cycle and their respective speeds. The average synchronization time cost per delivery point can be determined using the following formulation:

$$C_s = \frac{1}{\sqrt{3Q}} \left( \frac{\frac{2n}{n+1}\sqrt{k_{dt}^2 + \left(\frac{n+1}{2}\right)^2 \frac{1}{k_{dt}^2} \frac{n+1}{n}}}{V_d} - \frac{\left(\frac{k_{dt}}{n+1} + \frac{1}{k_{dt}}\right)(n+1)}{V_t} \right) c_w, \tag{6}$$

where $V_d$ is the drone tour speed in delivery area, $c_w$ is the value of waiting time.

The number of delivery points per route $m_{DT}$ in drone-assisted truck delivery can be computed similarly with Eq. (G4):

$$m_{DT} = \frac{T - \frac{L}{V_l}}{\frac{1}{V_t}\left(\frac{1}{n+1} * \frac{w_{dt}}{3} + \frac{1}{Q w_{dt}}\right) + \frac{1}{n+1} d_t}. \tag{7}$$

where the optimal swath width of expected drone-assisted truck delivery per delivery point $w^* \cong \sqrt{n+1}\sqrt{\frac{3}{Q}}\left(\frac{C_t^o + \sqrt{2} n C_d^o}{C_t^o + 2n C_d^o}\right)$, as approximated by Campbell et al. (2017).

Substituting the Eq. (7) into (10) leads to

$$E_{DT} = \frac{1}{\sqrt{3Q}} \left( C_t^o \left( \frac{k_{dt}}{n+1} + \frac{1}{k_{dt}} \right) + C_d^o \frac{2n}{n+1} \sqrt{k_{dt}^2 + \left(\frac{n+1}{2}\right)^2 \frac{1}{k_{dt}^2}} \right) + C_t^o \frac{L}{m_{DT}} + \frac{n}{n+1} S_d + S_t + C_s, \tag{8}$$

where $k_{dt} = \sqrt{n+1}\left(\frac{1+\sqrt{2}n\frac{C_d^o}{C_t^o}}{1+2n\frac{C_d^o}{C_t^o}}\right)$, $n$ is the number of drones per truck, $S_d$ is the marginal cost of drone delivery relative to truck delivery, which includes launch and recovery cost, $\frac{n}{n+1}$ is the fraction of stops served by drones.

The total cost of drone-assisted truck delivery in the service zone is:

$$C_{DT} = E_{DT} * A * Q = \alpha_2 Q + \beta_2 Q^{\frac{1}{2}}, \quad (9)$$

where $\alpha_2 = \left(\frac{LC_t^o d_t V_l}{(V_l T - L)(n+1)} + \frac{n}{n+1} S_d + S_t\right) A$ , $\beta_2 = \left(\left(1 + \frac{LV_l}{V_d(V_l T - L)} - \frac{(n+1)c_w}{V_t}\right)\left(\frac{k_{dt}}{n+1} + \frac{1}{k_{dt}}\right) + \left(\frac{2nC_d^o}{C_t^o(n+1)} + \frac{2c_w}{V_d}\right)\sqrt{{k_{dt}}^2 + \left(\frac{n+1}{2k_{dt}}\right)^2}\right)\frac{C_t^o}{\sqrt{3}} A.$

As shown in Eqs. (2) and (9), the cost function for both the truck-only service and the drone-assisted truck service can be formulated in terms of demand density. To establish a static switching model, an instantaneous cost-saving function $\Omega(Q)$ that compares the two services will be created by computing the cost difference between Eq. (6) and (16), which is shown in Eq. (17). Whenever this cost-saving becomes positive, a decision to switch is triggered. In the IC switching model, it is assumed that the decision to switch between service types occurs instantaneously or with negligible impact on the time horizon.

$$\Omega(Q) = C_{TO}(Q) - C_{DT}(Q) = \alpha_3 Q + \beta_3 Q^{\frac{1}{2}}, \quad (10)$$

where $\alpha_3 = \frac{n}{n+1}\left(\frac{L d_t V_l C_t^o}{V_l T - L} - S_d\right) A$, $\beta_3 = \left(\left(1 + \frac{LV_l}{V_d(V_l T - L)}\right)\left(2 - \left(\frac{k_{dt}}{n+1} + \frac{1}{k_{dt}}\right)\right) + \frac{(n+1)c_w}{V_t}\left(\frac{k_{dt}}{(n+1)} + \frac{1}{k_{dt}}\right) - \left(\frac{2nC_d^o}{C_t^o(n+1)} + \frac{2c_w}{V_d}\right)\sqrt{{k_{dt}}^2 + \left(\frac{n+1}{2k_{dt}}\right)^2}\right)\frac{C_t^o}{\sqrt{3}} A.$

Eq. (10) represents a demand density formula characterized by the parameters $\alpha_3$ and $\beta_3$, both of which are defined by given parameters in Table 1.

# 4. Stochastic Dynamic Switching Model

This section presents a stochastic modeling framework for demand-driven service switching decisions using the entry-exit real option approach. Demand density is modeled as a Geometric Brownian Motion (GBM) process to account for uncertainties influenced by various dynamic factors. The proposed approach establishes upper and lower demand thresholds that trigger service transitions, incorporating switching costs and hysteresis effects to determine optimal timing. While historical demand is known, estimating future demand poses a significant challenge. In a realistic setting, the decision to switch between the deployment of two services is heavily reliant on demand in the near future, which follows a stochastic process that can be quantified using the entry-exit real option approach discussed in section 2.2. In the context of the entry-exit real option framework, the utilization of GBM serves as a tool for modeling unpredictable fluctuations in demand. These fluctuations are influenced by dynamic factors such as economic conditions, demographic shifts, pandemics, and other potential variables. It is widely employed in financial modeling due to its ability to simulate the stochastic movements of financial assets (Stojkoski et al., 2020). Therefore, we assumed the demand density to be a time-dependent stochastic process, adhering to GBM, which can be represented by Eq. (11).

$$dQ(t) = \mu Q(t)dt + \sigma Q(t)dw(t), \tag{11}$$

where $\mu$ is the growth rate, $\sigma$ is the volatility of the process, and $dw$ is the increment of a standard Wiener process.

The solution to Eq. (11) is provided in Eq. (12) (Dixit, 1994).

$$Q(t) = Q(0)\exp\left(\left(\mu - \frac{1}{2}\sigma^2\right)t + \sigma w\right). \tag{12}$$

The expected value and conditional variance of $Q(t)$ are shown in Eqs. (13) and (14), respectively (Øksendal, 2003).

$$E_0[Q(t)] = Q(0)\exp(\mu t), \tag{13}$$

$$Var_0[Q(t)] = Q(0)^2\exp(2\mu t)\left(\exp(\sigma^2 t) - 1\right). \tag{14}$$

Various methods are available for estimating the parameters of the GBM model, including the maximum likelihood approach (Hurn et al., 2003), the fractional or Gaussian noise driven method (Ibrahim et al., 2021), and sequential Monte Carlo (Croghan et al., 2017).

The problem of stochastic switching between delivery services is addressed using the "asset equilibrium pricing", which involves building differential equations for the dynamic equilibrium value of switching based on a stochastic process (Guo et al., 2017). Let $V_0(Q)$ be the option value of operating a TO delivery service when demand density is $Q$, and $V_1(Q)$ be the option value of operating a DT delivery service. The asset equilibrium condition, as per Ito's lemma (Dixit, 1989), is equivalent to the second order differential equation for the TO delivery service described in Eq. (15. Similar computations can determine the option value for DT delivery service, with additional immediate cost saving $\Omega(Q)$ over TO delivery. The ordinary differential equation can be solved to obtain the value of $V_1(Q)$ that complies with the asset equilibrium condition, as expressed in Eq. (16):

$$\frac{1}{2}\sigma^2 Q^2 V_0''(Q) + \mu Q V_0'(Q) - \rho V_0(Q) = 0, \tag{15}$$

$$\frac{1}{2}\sigma^2 Q^2 V_1''(Q) + \mu Q V_1'(Q) - \rho V_1(Q) + \Omega(Q) = 0. \tag{16}$$

where $V_0(Q)$ is option value of using TO delivery service, $V_1(Q)$ is option value of using DT delivery service, $\Omega(Q)$ is given in Eq. (10), and $\rho$ is the discount rate over a time increment.

Due to the presence of switching costs, the single threshold created to determine the best switching timing is now divided into two thresholds, including upper and lower thresholds. Therefore, the demand thresholds $Q_H$ (from TO delivery to DT delivery) and $Q_L$ (from DT delivery to TO delivery), respectively, are what cause entry and exit for service switching. The following "value matching" interactions between the option values are created by these thresholds, as expressed in Eqs. (17) and (18).

$$V_0(Q_H) = V_1(Q_H) - F^+, \tag{17}$$

$$V_1(Q_L) = V_0(Q_L) - F^-. \tag{18}$$

where $F^+$ is the assumed cost for switching from TO delivery to DT delivery, $F^-$ is the assumed switching cost from DT delivery to TO delivery, $Q_H$ is the upper demand trigger when switching from TO delivery to DT delivery, and $Q_L$ is the lower demand trigger for switching from DT delivery to TO delivery.

To satisfy the "smooth pasting condition" (Dixit, 1989), the first order derivatives of Eqs. (19) and (20) can be expressed as:

$$V_0'(Q_H) = V_1'(Q_H), \tag{19}$$

$$V_0'(Q_L) = V_1'(Q_L). \tag{20}$$

When demand density $Q$ is assumed to grow according to the GBM model as in Eq. (11), we have a general solution of $V_0(Q)$ using infinite series (Dixit, 1989), as shown in Eq. (21):

$$V_0(Q) = A_0 Q^{\gamma_0} + B_0 Q^{\gamma_1}. \tag{21}$$

Similarly, for $V_1(Q)$ the solution to Eq. (16) is obtained in Eq. (22).

$$V_1(Q) = A_1 Q^{\gamma_0} + B_1 Q^{\gamma_1} - \frac{\alpha_3}{\mu-\rho} Q - \frac{\beta_3}{\frac{1}{2}\left(\mu - \frac{\sigma^2}{4}\right) - \rho} Q^{\frac{1}{2}}. \tag{22}$$

where $A_0$ , $A_1$, $B_0$, $B_1$ are constants to be determined, while $\gamma_0$, $\gamma_1$ are the negative and positive root of the quadratic equation in Eq. (23):

$$\frac{1}{2}\sigma^2\gamma(\gamma - 1) + \mu\gamma - \rho = 0, \tag{23}$$

Appendix A provides the proof of the expression of $V_1(Q)$ from Eq. (22).

**Proposition 1.** The upper and lower demand thresholds can be derived by solving the following system of two non-linear equations:

$$Q_H^{-\gamma_1} \frac{(1-\gamma_0)K_1(Q_H)+\left(\frac{1}{2}-\gamma_0\right)K_2(Q_H)+\gamma_0 F^+}{\gamma_1-\gamma_0} = Q_L^{-\gamma_1} \frac{(1-\gamma_0)K_1(Q_L)+\left(\frac{1}{2}-\gamma_0\right)K_2(Q_L)-\gamma_0 F^-}{\gamma_1-\gamma_0}, \quad (24)$$

$$Q_H^{-\gamma_0} \frac{(1-\gamma_1)K_1(Q_H)+\left(\frac{1}{2}-\gamma_1\right)K_2(Q_H)+\gamma_1 F^+}{\gamma_1-\gamma_0} = Q_L^{-\gamma_0} \frac{(1-\gamma_1)K_1(Q_L)+\left(\frac{1}{2}-\gamma_1\right)K_2(Q_L)-\gamma_1 F^-}{\gamma_1-\gamma_0}. \quad (25)$$

From Eqs. (24) and (25) we can obtain $Q_H$ and $Q_L$ numerically and then determine $A_1$ and $B_0$.

**Proof.** Proof of Proposition 1 is provided in Appendix B.

The definitions of the demand thresholds, $Q_H$ and $Q_L$, originate from the "hysteresis effect" concept proposed by Dixit (1989). The hysteresis effect, attributed to the switching costs, gives rise to a state of "indifference band" in the demand density that creates the lower and upper thresholds, which will be discussed in the sensitivity analysis section. While deriving analytical solutions for $Q_H$ and $Q_L$ presents a challenge, Proposition 2 facilitates the analytical determination of the expected transition time between delivery services.

**Proposition 2.** For $Q(0) = Q < Q_H$, the expected duration for transition from the truck-only delivery service to the drone-assisted truck delivery service can be expressed as follows:

$$E_{Q_L}[\tau_H] = \frac{\ln Q_H}{\eta-\frac{\sigma^2}{2}}\left(\frac{Q_L}{Q_H}\right)^l - \frac{\ln Q_L}{\eta-\frac{\sigma^2}{2}}. \quad (26)$$

Similarly, for $Q(0) = Q < Q_L$, the expected duration for transition from the drone-assisted truck delivery service to the truck-only delivery service can be expressed as follows:

$$E_{Q_H}[\tau_L] = \frac{\ln Q_L}{\eta-\frac{\sigma^2}{2}}\left(\frac{Q_H}{Q_L}\right)^l - \frac{\ln Q_H}{\eta-\frac{\sigma^2}{2}}. \quad (27)$$

where $l$ is a parameter which can be determined by $l = 1 - \frac{2\eta}{\sigma^2}$.

**Proof.** Proof of Proposition 2 is provided in Appendix C.

In a particular scenario, when the switching cost, $F^+$ or $(F^-)$, s reduced to zero, the two switching thresholds, $Q_H$ and $Q_L$, merge into a single threshold denoted as $Q^*$. Given this situation, the equations represented by Eq. (17-20) can be reformulated as follows:

$$V_0(Q^*) = V_1(Q^*), \quad (28)$$

$$V_0'(Q^*) = V_1'(Q^*). \quad (29)$$

To find $Q^*$, the following additional condition is needed:

$$V_0''(Q^*) = V_1''(Q^*). \tag{30}$$

**Proposition 3.** The single threshold $Q^*$ when the switching cost $F^+(or\ F^-)$ is zero, can be found by solving the following non-linear equation:

$$\gamma_1(\gamma_1 - 1)\frac{(1-\gamma_0)K_1(Q^*)+\left(\frac{1}{2}-\gamma_0\right)K_2(Q^*)}{\gamma_1-\gamma_0} - \gamma_0(\gamma_0 - 1)\frac{(1-\gamma_1)K_1(Q^*)+\left(\frac{1}{2}-\gamma_1\right)K_2(Q^*)}{\gamma_1-\gamma_0} + \frac{K_2(Q^*)}{4} = 0. \tag{31}$$

**Proof.** Proof of Proposition 3 is provided in Appendix D.

In the absence of a switching cost in the stochastic model, the conditions of "value matching" and "smooth pasting" from Eqs. (28-30) can be substituted into Eqs. (15) and (16), This results in the instantaneous cost-saving $\Omega(Q)$ becoming zero. Consequently, the single threshold $Q^*$ degrades to the deterministic threshold (Dixit and Pindyck, 1994), and will be verified in the model application section.

This section illustrates the application of our models to evaluate the cost-effectiveness of truck-only (TO) and drone-assisted truck (DT) delivery services under varying demand conditions. We first present a static analysis to understand the relationship between cost savings and demand density. This analysis reveals the demand thresholds at which switching between TO and DT services becomes economically advantageous.

Subsequently, a dynamic analysis that incorporates demand uncertainty in conducted, modeled as a GBM process. This allows us to determine optimal switching strategies that minimize costs in a dynamic environment. By comparing the results of the immediate cost-saving, deterministic, and stochastic models, we demonstrate the importance of considering switching costs and demand uncertainty when making operational decisions. The findings highlight the potential for significant cost savings and improved efficiency through the use of dynamic switching strategies.

# 5. Model Application

## 5.1. Static model analyses

Unless otherwise noted, the input parameters and baseline values in Table 1 are mostly from Choi and Schonfeld (2021) or Baldisseri et al. (2022). Other values of some key parameters are used in sensitivity analyses in Table 2 from Section 5.1.

**Table 2.** The parameters in dynamic models.

| Parameters | Definitions | Baseline values | Parameter range | Units |
|---|---|---|---|---|
| $A$ | Service zone | 1250 | -- | sq mi |
| $C_t^o, C_d^o$ | Operation cost of truck and drone per unit distance | | | |
| $d_t$ | Average stop time per truck delivery at a delivery point | | | |
| $n$ | Number of drones per truck | 10 | -- | -- |
| $S_t$ | Fixed truck stop cost | 0.2 | -- | $ |
| $S_d$ | Marginal cost for drone delivery, including launch and recovery cost, relative to truck delivery | 0 | -- | $ |
| $T$ | Expected route time | 8 | 4-12 | hrs |
| $V_l$ | The truck linehaul speed | 60 | 40-80 | mi/hr |
| $V_t, V_d$ | The truck and drone tour speed in delivery area | 30 | 15-45 | mi/hr |
| $\phi$ | Circuity factor for line-haul distance | 2/3 | | |
| $\nu$ | Adjust factor for line-haul distance | 1.2 | | |
| $Q(0)$ | Demand density at time 0 | 50 | -- | pkg/ sq mi/day |
| $H_T$ | Horizon planning time | 12 | -- | Months |
| $\rho$ | Discount rate | 0.025 | 0.01-0.04 | -- |
| $\mu$ | Demand growth rate | 0.5% | 0.2-0.8% | -- |
| $\sigma$ | Volatility rate | 10% | 5%-15% | -- |
| $F^+(F^-)$ | Switching cost | 1000 | 500-1500 | $ |
| $Z$ | Area ratios of the region and delivery zone | 5 | | Times |
| $\varepsilon_{TO}, \varepsilon_{DT}^n$ | Proportion of customer points delivered by TO and DT(n-drones) | 0.25 | | |

First, the relation between the saved cost and the service zone's demand density is computed. The saved cost is the difference between the cost of TO delivery and that of DT. A positive saved cost indicates that truck-only delivery is more expensive than drone-assisted delivery and vice versa. According to Figure 5, TO delivery is more cost-effective than DT when the demand density is on the lower end, say, under roughly 70.2 packages/mi$^2$ per day. Therefore, we can see that switching to DT delivery at low demand density causes losses to the delivery system. As demand density increases, the rate of cost saved also accelerates. However, this static model does not factor in the cost of switching, even if the switching cost exceeds the accumulated saved cost at that moment. Consequently, the transition from TO to DT occurs as soon as the cost of DT drops below that of TO.

Furthermore, Figure 6 illustrates the influence of the number of drones per truck on the overall cost. The graph demonstrates that additional drones per truck delay the point of intersection between these services. The increase in the number of drones creates a clustering effect from economies of scale, expanding the benefits, which leads to increased saved cost between two services as the number of drones increases.

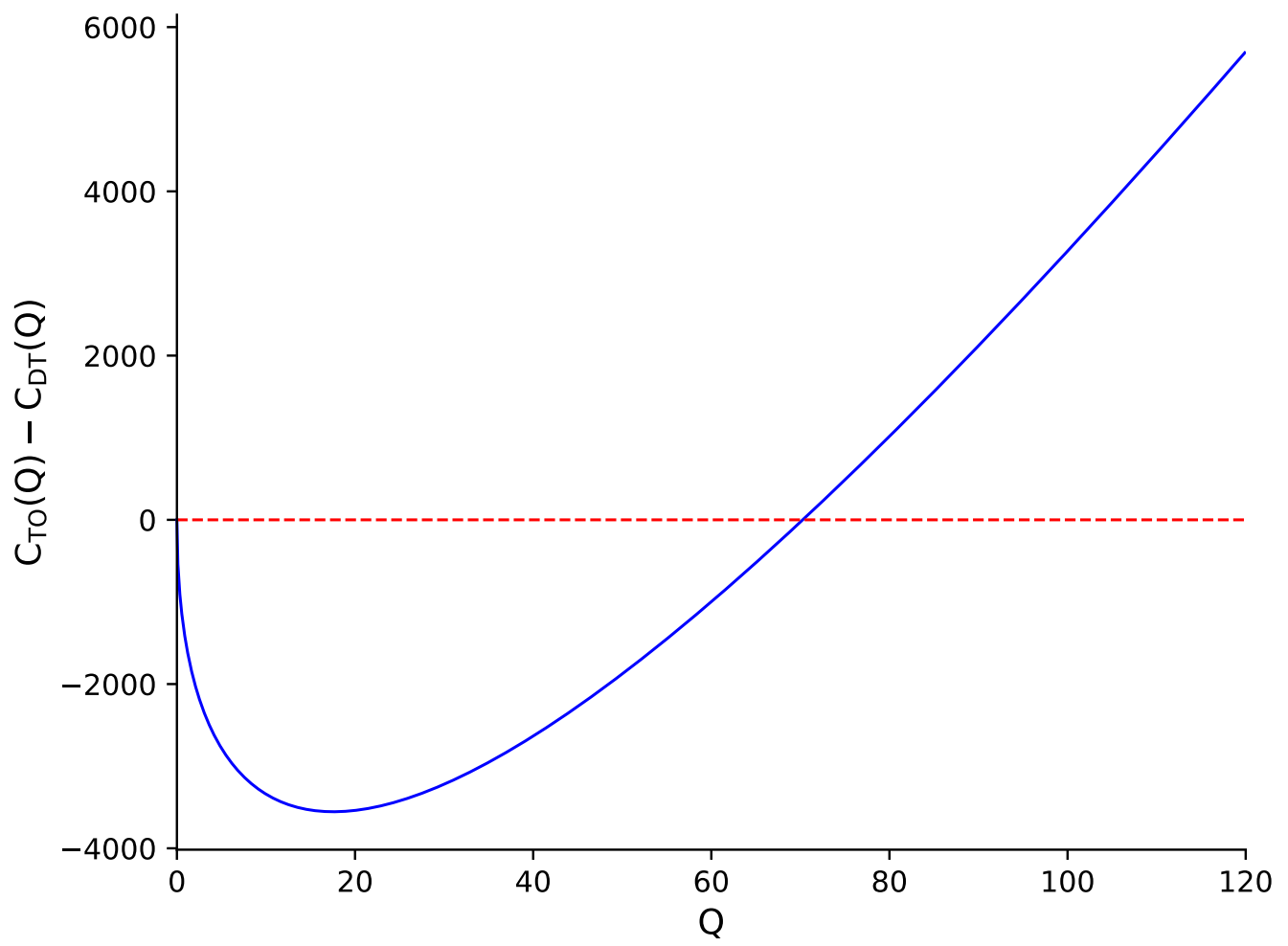

**Figure 5.** Saved cost of truck-only compared to drone-assisted truck delivery.

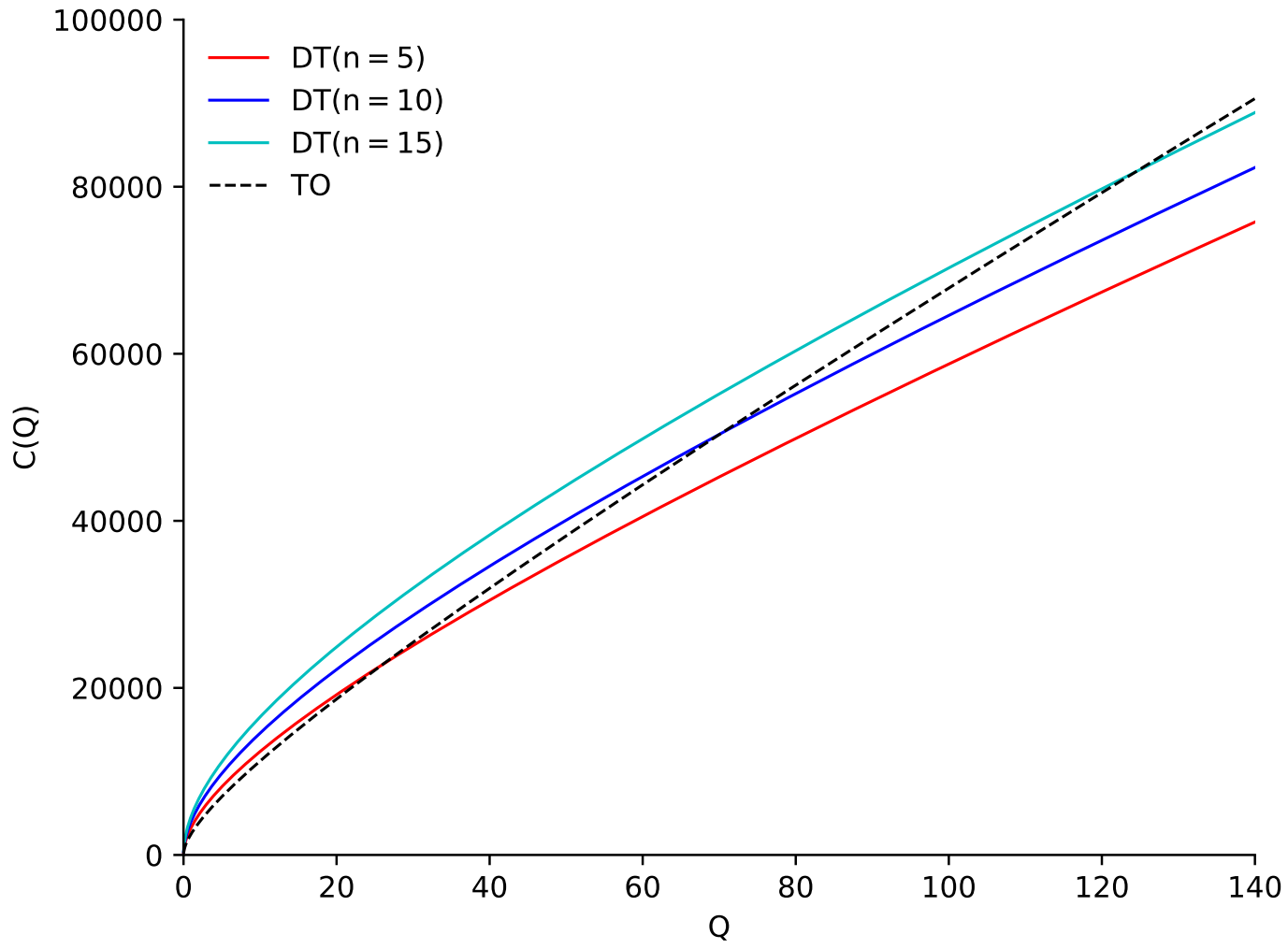

**Figure 6.** Sensitivity of total cost to the number of drones per truck.

Next, we delve into the sensitivity analysis of saved cost, focusing on several key parameters: route time, average truck dwell time at stops, truck operating cost, drone operating cost, line haul speed by truck, and local/tour speed by truck. Table 1 illustrates the values for each parameter, including one value above the baseline and another below it. In this sensitivity analysis, only one parameter at a time is modified while all other parameters remain unchanged at their baseline values.

The outcomes of this examination reveal that, given a specific demand density, the time-related saved cost increases as the time value rises. As depicted in Figure 7 (a), the growth rate of saved cost diminishes with the increase in route time T, consequently deferring the switching point. Conversely, when the average truck dwell time decreases, the rate of saved cost growth decelerates, which postpones the switching point, as seen in Figure 7 (b).

From Figures 7 (c) and (d), it becomes apparent that truck operating costs and drone operating costs exhibit divergent trends, both impacting the overall saved cost of the delivery system. A decrease in drone operating costs $C_d^o$ accelerates the introduction of drone-assisted trucks, whereas a decrease in truck operating costs $C_t^o$ delays the adoption of the truck-only delivery service. Furthermore, increasing either the truck's line haul speed or its local speed reduces the cost of truck-only delivery, yet both actions delay the switching point between these two operational services, as seen in Figures 7 (e) and (f).

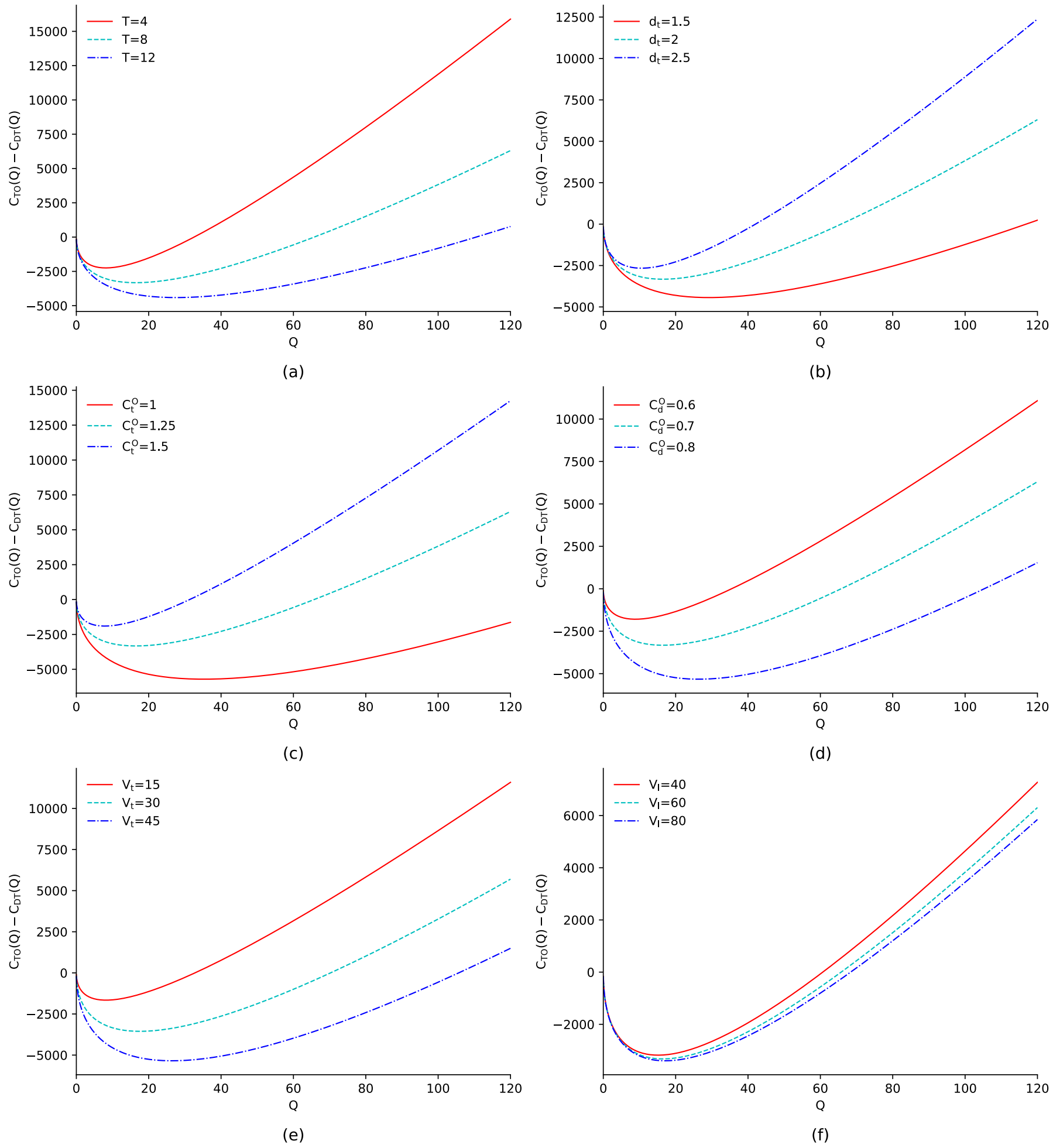

**Figure 7.** Sensitivity of saved cost to various parameters.

## 5.2. Dynamic switching analyses

As discussed earlier, GBM effectively captures the dynamic and time-sensitive nature of demand uncertainty. Consequently, in this section, we investigate a scenario where the demand density follows a GBM trajectory. We use the parameters outlined in Table 2 to simulate the daily demand in one year depicted in Figure 8.

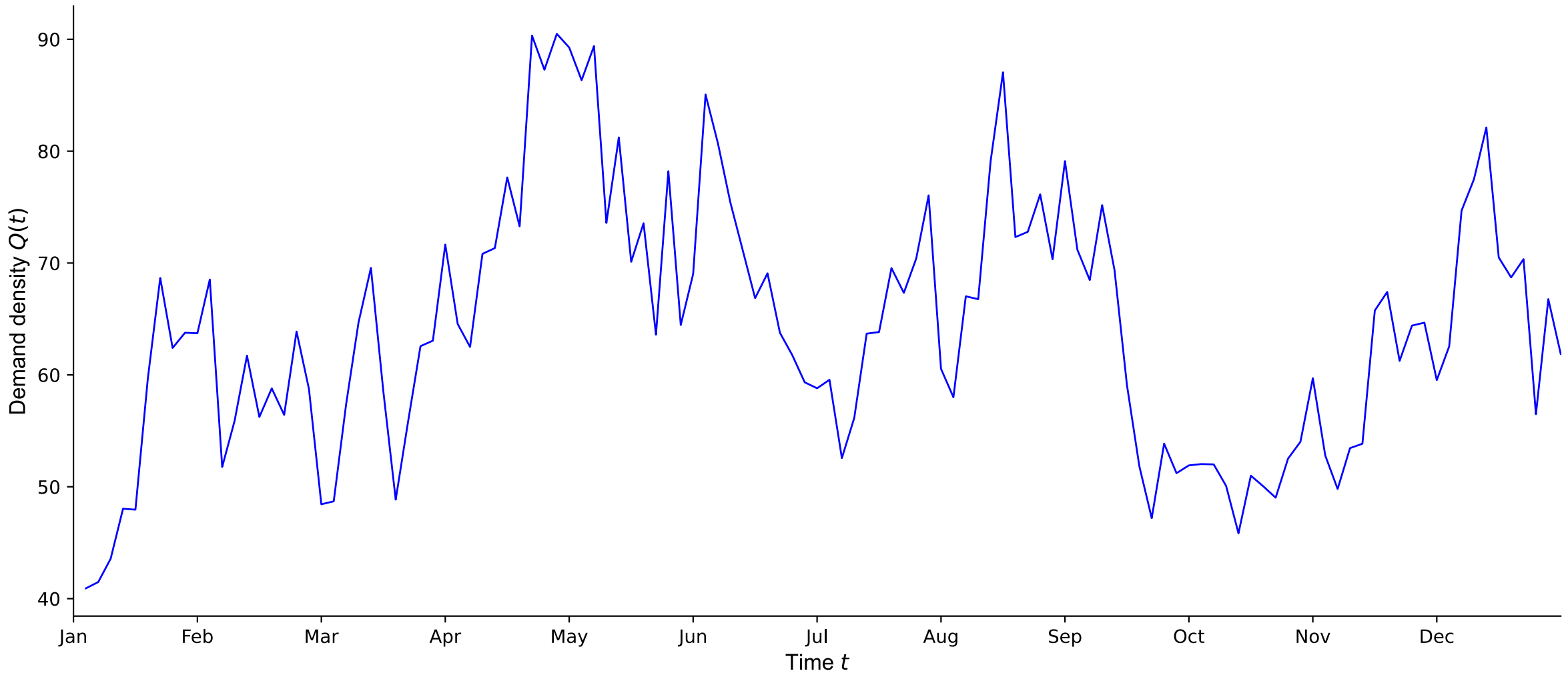

**Figure 8.** Exogenous trajectory of demand density over time.

### 5.2.1. Immediate cost-saving switching model analyses

In our analysis of dynamic switching, we explore three models: the immediate cost-saving (IC) switching model, the deterministic dynamic switching model, and the dynamic stochastic switching model. The IC switching model is characterized by the operator switching as soon as saved cost exceed zero. Whenever the initial saved cost between the two delivery services exceeds 0, the system automatically switches to the service with the lower cost.

$$C_{TO} - C_{DT} \geq 0, \quad (32)$$

The overall cost for both TO and DT fluctuates due to the dynamic evolution of the demand density in Figure 9. The instances of transition between the two services are indicated by the intersections of their respective cost curves on the graph. In the IC model, when the total cost of DT delivery falls below that TO delivery, the system switches from TO to DT, and vice versa. These transition points are highlighted with green circles. As illustrated in Figure 9, during the initial planning phase, truck-only delivery is the preferred service due to its lower cost. However, when the cost of DT delivery drops below that of TO delivery, a switch occurs.

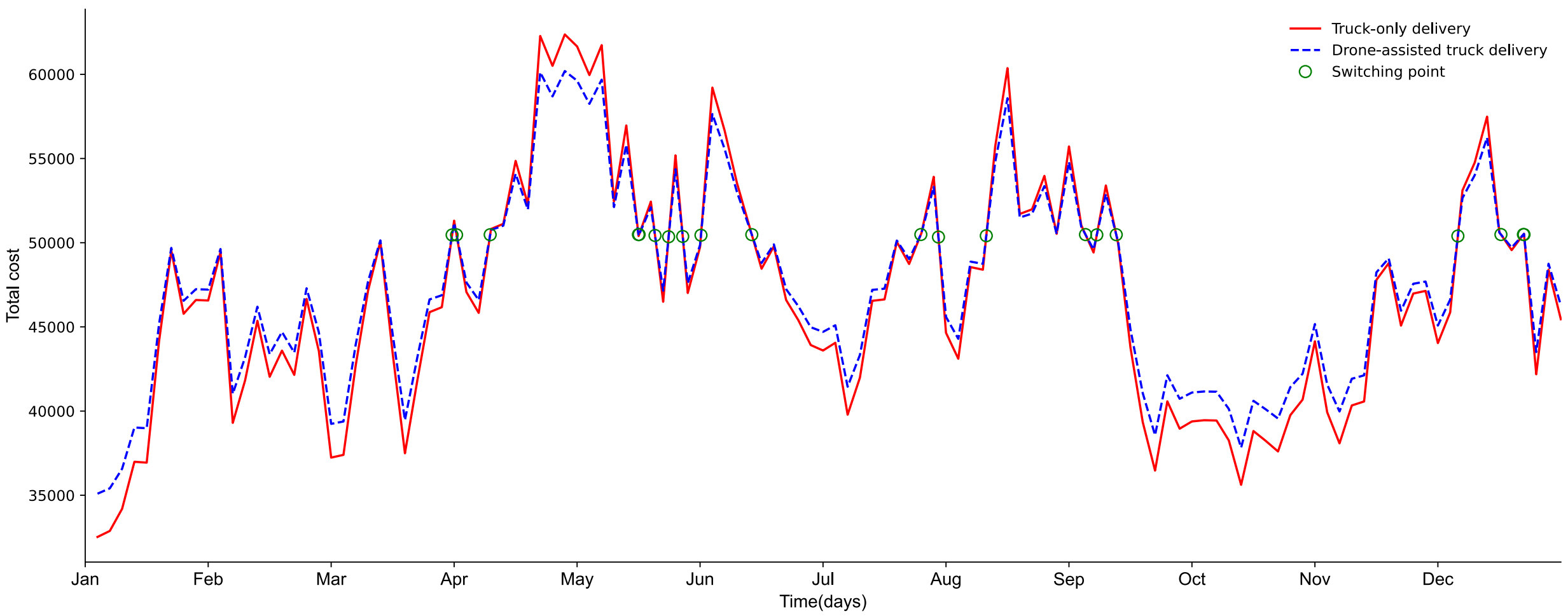

**Figure 9.** Total cost under different delivery service.

### 5.2.2. Deterministic switching model analyses

In the deterministic switching model, switching occurs when the cost difference between the two services surpasses the switching cost threshold. If the switching cost, $F$, is considered whenever a service switch occurs, the decision to switch delivery services is based on whether the initial saved cost between the two delivery services exceeds the switching cost, denoted as $F$. Each of these transitions from TO to DT or vice versa entails a one-time switching cost of $F$. In this case, the system assesses whether the initial cost difference reaches or exceeds the switching costs to ensure that each switching instance is profitable.

$$C_{TO} - C_{DT} \geq F, \tag{33}$$

This transition leads to saved cost over time, which can be evaluated according to Equation (33) before the shift from TO to DT is initiated. As shown in Figure 10, the initial service is TO, and when the cost difference exceeds *0*, using the DT service becomes preferrable to using TO and vice versa. This figure also shows the effect of considering switching cost when determining the switching point in the deterministic switching model. As also shown in Figure 10, when switching cost is not considered, the switching point is represented by black circles and triangles, and the system switches from TO to DT whenever the saved cost exceeds 0, and vice versa for switching from DT to TO. Incorporating the switching cost leads to a delay in the switching timing, as each switch must provide a greater saved cost than the switching cost to justify the transition. As switching costs increase, the frequency of switches decreases.

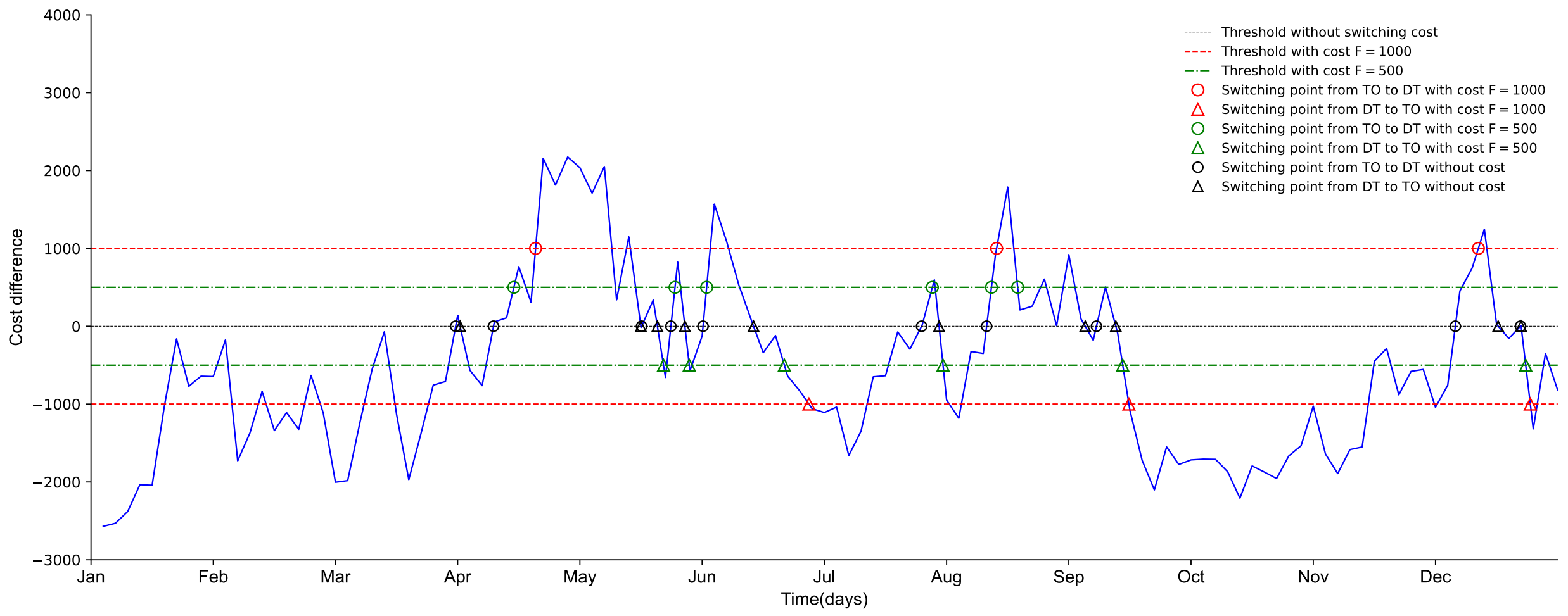

**Figure 10.** Switching points with different switching costs.

### 5.2.3. Stochastic switching model analyses

In the proposed stochastic model, switching thresholds are derived in accordance with section 4. By establishing these switching thresholds for the three model types, we can determine the switching timings under various scenarios.

Simulations of the demand density fluctuations using the GBM model over a span of 12 months (each unit representing one day) are conducted. The initial service of service is set as truck-only delivery due to the lower demand density. Figure 11 illustrates the simulated trajectory of the demand density over time and the switching time point in a scenario where switching costs are not considered. In this scenario with zero switching cost, there is only one threshold, $Q^* = 70.6$. When the demand density exceeds that single threshold, the delivery system switches from TO to DT, which is indicated by the red circles in Figure 11. Conversely, when the demand density falls below the same threshold, the system reverts to the TO service to optimize cost, as represented by the green circles in Figure 11.

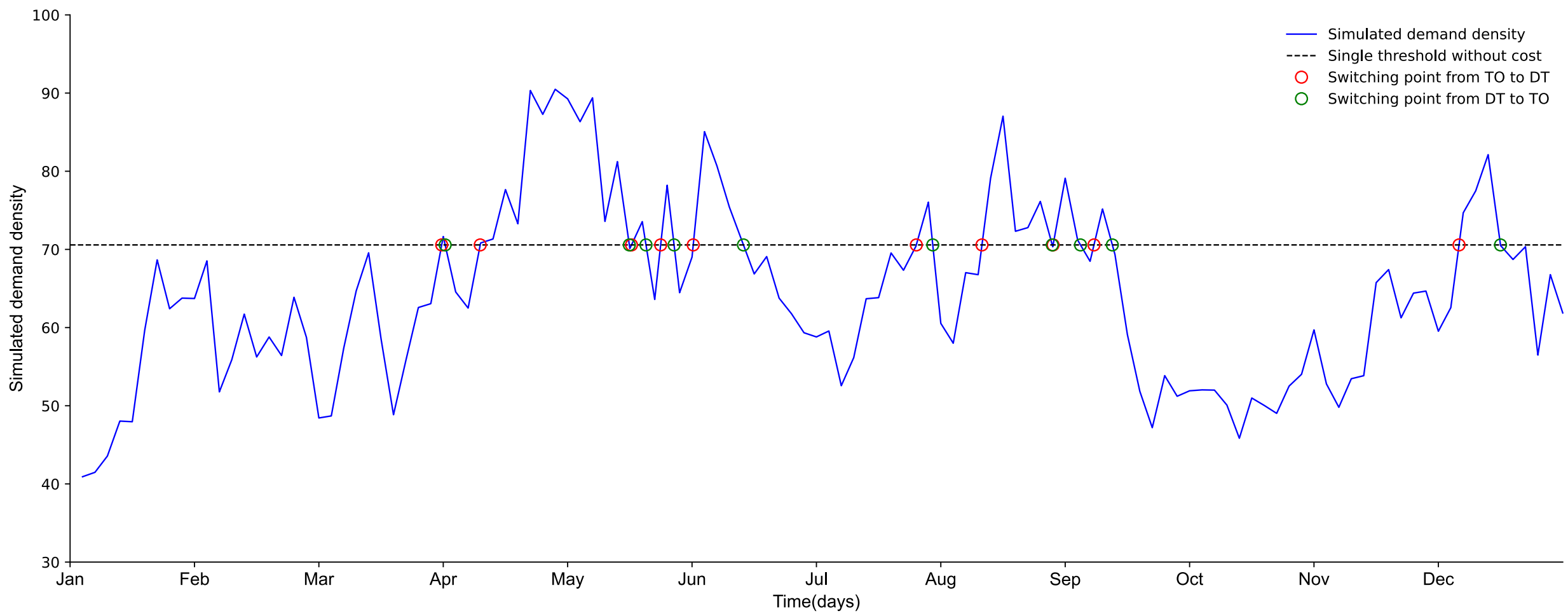

**Figure 11.** Switching timing of stochastic switching without cost.

In scenarios involving the consideration of switching costs, the stochastic switching model may identify two thresholds with Equations (24) and (25). The optimal thresholds are $Q_L = 63.8$ and $Q_H = 79.7$. Following this model, the switching time is postponed due to the necessity of accounting for switching costs. Consequently, the number of switching points is notably reduced compared to the stochastic model that does not account for switching costs, as shown in Figure 12.

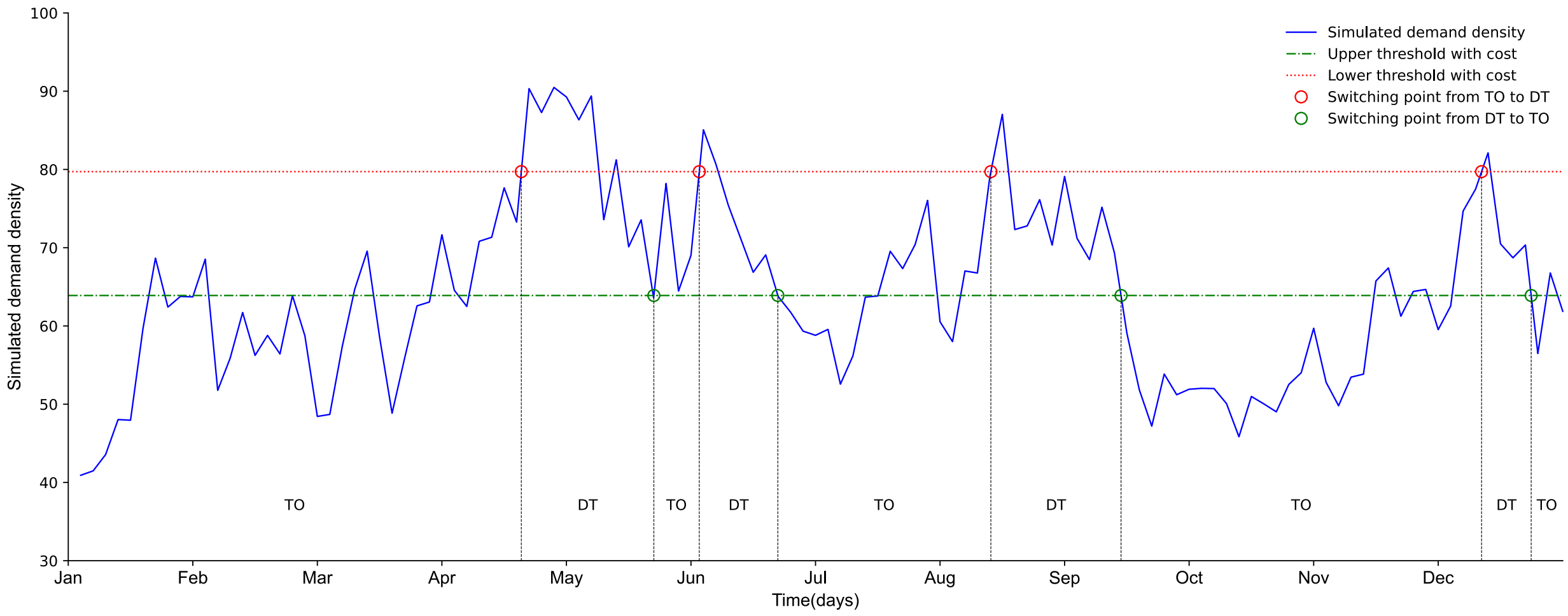

**Figure 12.** Switching timing of stochastic model with switching cost.

In the analysis of the three models, all drone-assisted trucks are assisted by 10 drones departing from the depot to the delivery area. Figure 9 indicates that switching occurs 20 times between the two services in the IC model, where the switching cost is not considered. However, when incorporating the switching cost in the deterministic model, as shown in Figure 10, the number of switching times decreases to 6 during the planning horizon. Some switches are excluded because the initial saved cost is below the switching cost. In contrast, our proposed dynamic stochastic model, illustrated in Figure 12, identifies 8 switching timings with the same switching cost as the deterministic model.

A comparison of the performances of various switching models is presented in Table 4. To compute the total saved cost, the saved cost in Equation (10) is integrated over the planning horizon. It can be observed that the saved cost achieved by the stochastic model increase substantially in contrast to the IC model. Specifically, this saved cost is 31.3% higher within the same planning horizon.

**Table 3.** Comparative analysis of diverse switching models' performance.

| | IC model | Deterministic model | Stochastic model |
|---|---|---|---|
| Total saved cost ($) | 61098.32 | 71729.42 | 80221.67 |
| (% from IC model) | 0 | 17.4% | 31.3% |

Table 4 presents the sensitivity of switching thresholds to various stochastic parameters and switching costs. Adjustments to the discount rate are made to accommodate shifts in both the upper and lower thresholds, as well as the single switching threshold. Notably, the distances from the single threshold are not symmetrical. An increase in the discount factor $\rho$ results in a decrease in $Q_L$ and an increase in $Q_H$. The entry-timing and exit-timing will be postponed in this situation. As volatility increases, the upper threshold $Q_H$ experiences a rise, while the lower threshold $Q_L$ declines. The indifference band widens as volatility rises. This outcome makes intuitive sense because greater uncertainty would postpone the entry-timing and exit-timing.

Table 4 also highlights that the demand density for the higher threshold increases, and the lower threshold decreases as switching costs increase from $500 to $1500. This observation underscores that the hysteresis effect emerges only in the presence of switching costs and grows more pronounced with higher switching costs. Consequently, a more advantageous strategy involves exercising patience and delaying action as switching costs increase.

Interestingly, the value of $Q^*$ remains unaffected by variations in stochastic parameters, as it converges to the deterministic threshold, as described by Dixit and Pindyck (1994). Hence, the stochastic parameters exhibit no influence on the optimal switching timing.

**Table 4.** Relations between various parameters and demand density thresholds.

| Parameter | Value | $Q^*$ | $Q_L$ | $(Q_L - Q^*)$ | $Q_H$ | $(Q_H - Q^*)$ |
|---|---|---|---|---|---|---|
| Discount rate $\rho$ | 0.01 | 70.6 | 69.52 | -1.08 | 75.46 | 4.86 |
| | 0.025 | 70.6 | 63.54 | -7.06 | 79.26 | 8.66 |
| | 0.04 | 70.6 | 57.38 | -13.22 | 82.84 | 12.24 |
| Demand growth rate $\mu$ | 0.002 | 70.6 | 67.62 | -2.98 | 79.14 | 8.54 |
| | 0.005 | 70.6 | 63.54 | -7.06 | 78.89 | 8.29 |
| | 0.008 | 70.6 | 60.66 | -9.94 | 76.92 | 6.32 |
| Volatility rate $\sigma$ | 0.05 | 70.6 | 69.11 | -1.49 | 73.85 | 3.25 |
| | 0.1 | 70.6 | 67.84 | -2.76 | 75.68 | 5.08 |
| | 0.15 | 70.6 | 65.79 | -4.81 | 78.23 | 7.63 |
| Switching cost $F^+(F^-)$ | 500 | 70.6 | 66.88 | -3.72 | 78.23 | 7.63 |
| | 1000 | 70.6 | 66.64 | -3.96 | 79.37 | 8.77 |
| | 1500 | 70.6 | 62.92 | -7.68 | 82.06 | 11.46 |

#### 5.2.4. Multiple options switching model analyses

The multiple option switching model extends beyond a simple binary choice by accommodating several distinct operational policies across various service zones, making it particularly well-suited for extensive regions with heterogeneous demand. Rather than mandating an immediate, region-wide shift from a truck-only (TO) fleet to a full drone-assisted truck (DT) deployment, this framework proposes a hybrid fleet (HD) configuration as a key transitional phase. This HD approach facilitates a more strategic and gradual tiered rollout of drone technology. Under an HD strategy, different service zones can concurrently deploy traditional TO vehicles alongside DT fleets, with the latter potentially equipped with varying numbers of drones. This allows for fine-grained, localized control over service capacity and operational costs, precisely tailored to the specific conditions of each zone. Even when only aggregate demand density data is available, the framework still guides the prioritization of initial DT introductions into the most concentrated demand areas, followed by a staged expansion into other zones as predefined demand thresholds are met.

Compared to an abrupt, one-step system-wide conversion between TO and DT operations, this multiple-option strategy, leveraging hybrid fleet deployments, ensures smoother operational transitions, minimizes potential service disruptions, and fosters a more cost-effective adaptation to evolving and spatially diverse demand patterns. The three resulting operational scenarios for the whole region, derived from this model, are described in detail in Table 5.

**Table 5.** Different scenarios for larger delivery regions

| Scenario | Full-region TO | HD | Full-region DT |
|---|---|---|---|
| Description | Region is served exclusively by TO fleet. | Region is served by different fleets including TO, DT(2-drones), | Region is served entirely using DT fleet. |
| | | DT(5-drones) and DT (10-drones). | |
| Operation illustration | 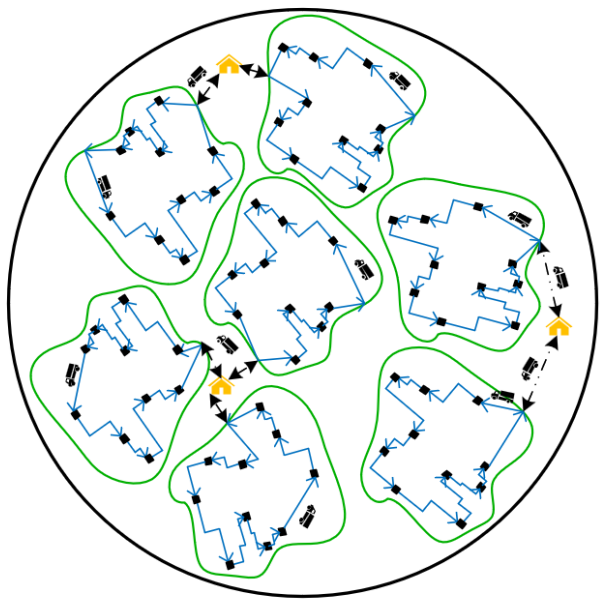 | 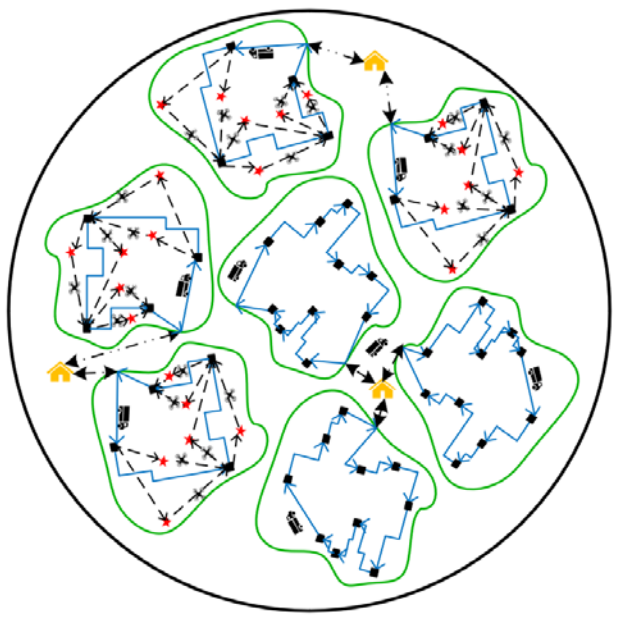 | 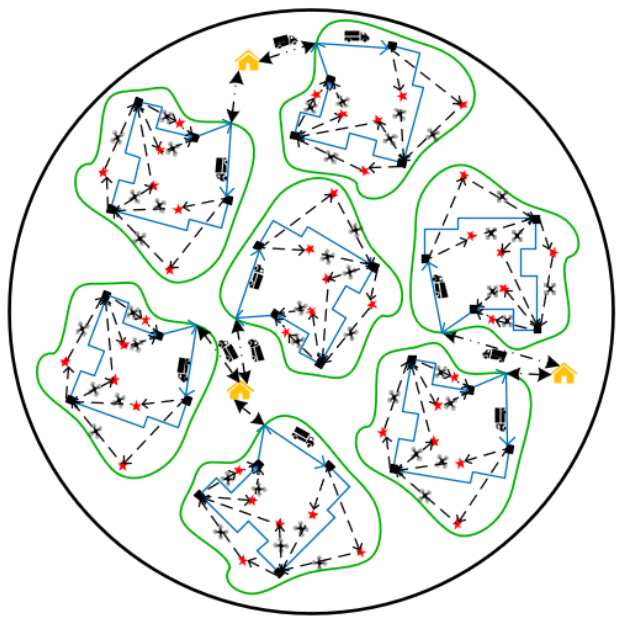 |
| Cost function | $P_{TO} = C_{TO}(ZQ)$ | $P_{HD} = C_{TO}(\varepsilon_{TO}ZQ) + \sum C_{DT}^{n}(\varepsilon_{DT}^{n}ZQ)$, | $P_{DT} = C_{DT}^{n}(ZQ)$ |
| | where $P$ is the cost to serve all delivery demand under different scenarios, $Z$ is area ratios of the region and delivery zone, $\varepsilon_{TO}$ and $\varepsilon_{DT}^{n}$ are the proportion of customer points delivered by TO and DT(n-drones). | | |

Figure 13 illustrates how costs vary, which may represent demand density under different scenarios (TO, DT, and HD). The three curves represent the costs associated with each scenario, all of which increase as demand density increases. Under static analyses, the scenario with the lowest cost changes depending on the value of demand density. For lower demand density (below 58.48), the TO has the lowest cost. When demand density is between 58.48 and 82.24, the HD becomes the optimal choice. For demand density above 82.84, the DT offers the lowest cost.

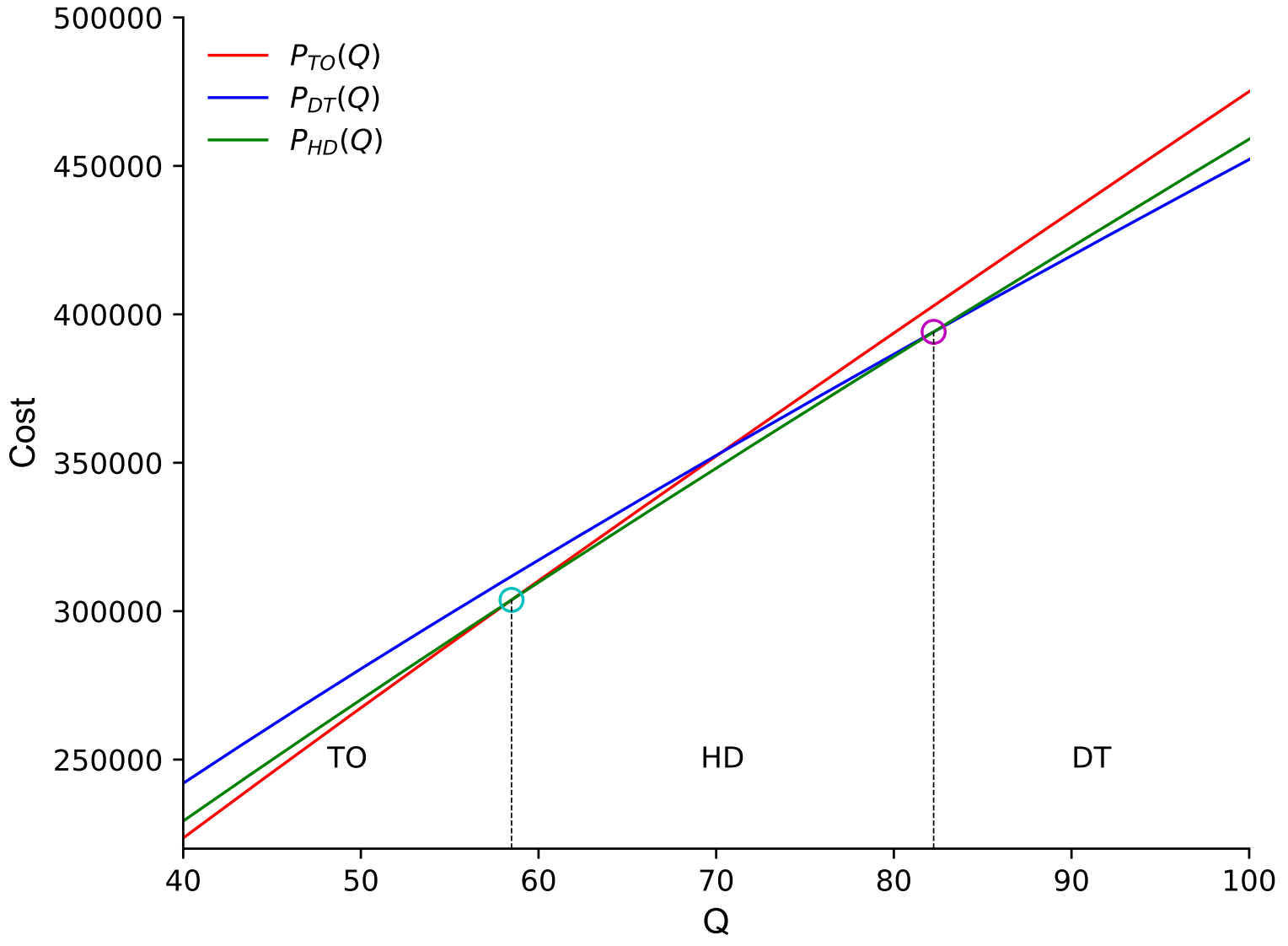

Figure 13. Total cost of different scenarios

In the multiple-option switching framework, the delivery system dynamically transitions through three primary operational states: initially from a truck-only (TO) configuration to a hybrid drone-assisted fleet (HD), and subsequently from the HD state to a full drone-assisted truck fleet (DT). The decision to transition between these states at each stage is determined by comparing mode-specific cost functions to compute the net benefit of switching. Under an infinite-horizon assumption and a specified stochastic demand model, optimal upper (activation) and lower (deactivation) switching thresholds for both stages are derived analytically (as presented in Equations (22)-(27) and summarized in Table 6). This results in four critical demand thresholds: two upper thresholds governing the transitions TO→HD and HD→DT, and two corresponding lower thresholds for the reverse transitions, DT→HD and HD→TO.

Figure 14 illustrates this process under simulated demand fluctuations. The system starts in the TO state. When demand exceeds the first upper threshold (yellow dashed line), it switches to HD (yellow dot). If demand continues rising beyond the second upper threshold (green dashed line), the system advances to DT (green dot). On the downside, a drop below the first lower threshold (red dashed line) triggers a return from DT to HD (red dot), and a further decline past the second lower threshold (magenta dashed line) reverts the system from HD back to TO (magenta dot). This two-stage, four-threshold policy ensures that the fleet configuration dynamically adapts to evolving demand, balancing cost efficiency with service capacity.

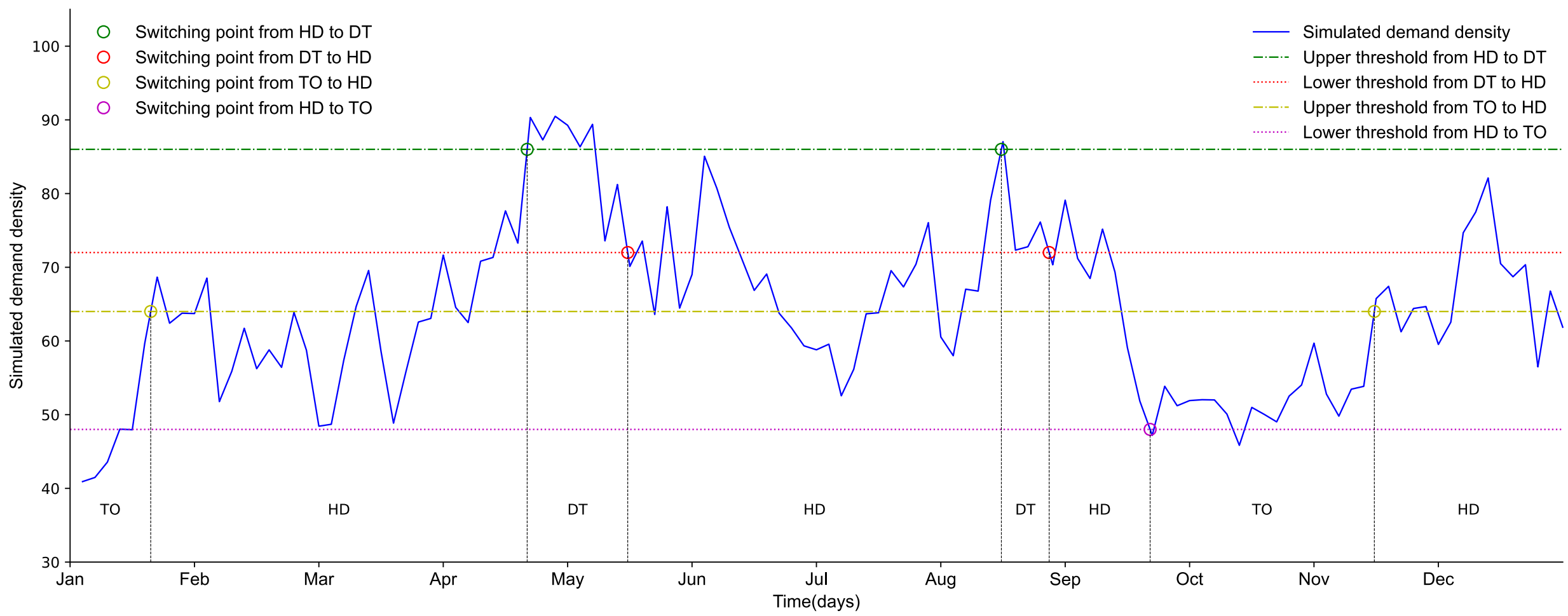

**Figure 14.** Switching timing of multiple options switching model with switching cost

Table 6 presents a comparative analysis of different delivery policies, contrasting static operational strategies with the proposed dynamic switching models. The policies under comparison include a baseline policy (Only TO) and another policy (Only DT). These are compared against two dynamic policies: a stochastic model that allows switching between TO and DT, and the more sophisticated multiple options model which introduces a hybrid (HD) phase, facilitating staged transitions from TO to HD and then from HD to DT. By examining the total cost savings of each policy relative to the truck-only baseline, the results indicate that statically deploying the drone-assisted system incurs higher costs. Conversely, dynamic switching policy demonstrate clear cost advantages. Notably, the multiple-options model, which permits phased transitions involving multiple fleet configurations, achieves the most substantial cost savings, highlighting its superior economic efficiency among the evaluated policies.

**Table 6.** Comparative analysis of different policies.

| | **Static policy** | | **Dynamic policy** | | | | |
|---|---|---|---|---|---|---|---|
| | | | | **Multiple options model (TO-HD-DT)** | | | |
| | **Truck-only (TO)** | **Drone assisted truck (DT)** | **Stochastic model (TO-DT)** | Stage 1 (TO-HD) | | Stage 2 (HD-DT) | |
| | | | | $Q_L^{TO-HD}$ | $Q_H^{TO-HD}$ | $Q_L^{HD-DT}$ | $Q_H^{HD-DT}$ |
| | | | | 49.1 | 64.3 | 72.1 | 86.3 |
| Total saved cost ($) compared with TO | 0 | -243746.95 | 119736.48 | 1486514.31 | | | |

## 5.3 Case study

Miami-Dade County (MDC), Florida is selected as the case study, with the base year for the dynamic switching model set as 2021. According to the census (U.S. Census Bureau, 2021a), approximately 2,663,000 individuals resided in MDC in 2021. On average, 65 package deliveries were generated per person in the U.S. diromg that year (Pitney Bowes 2022).

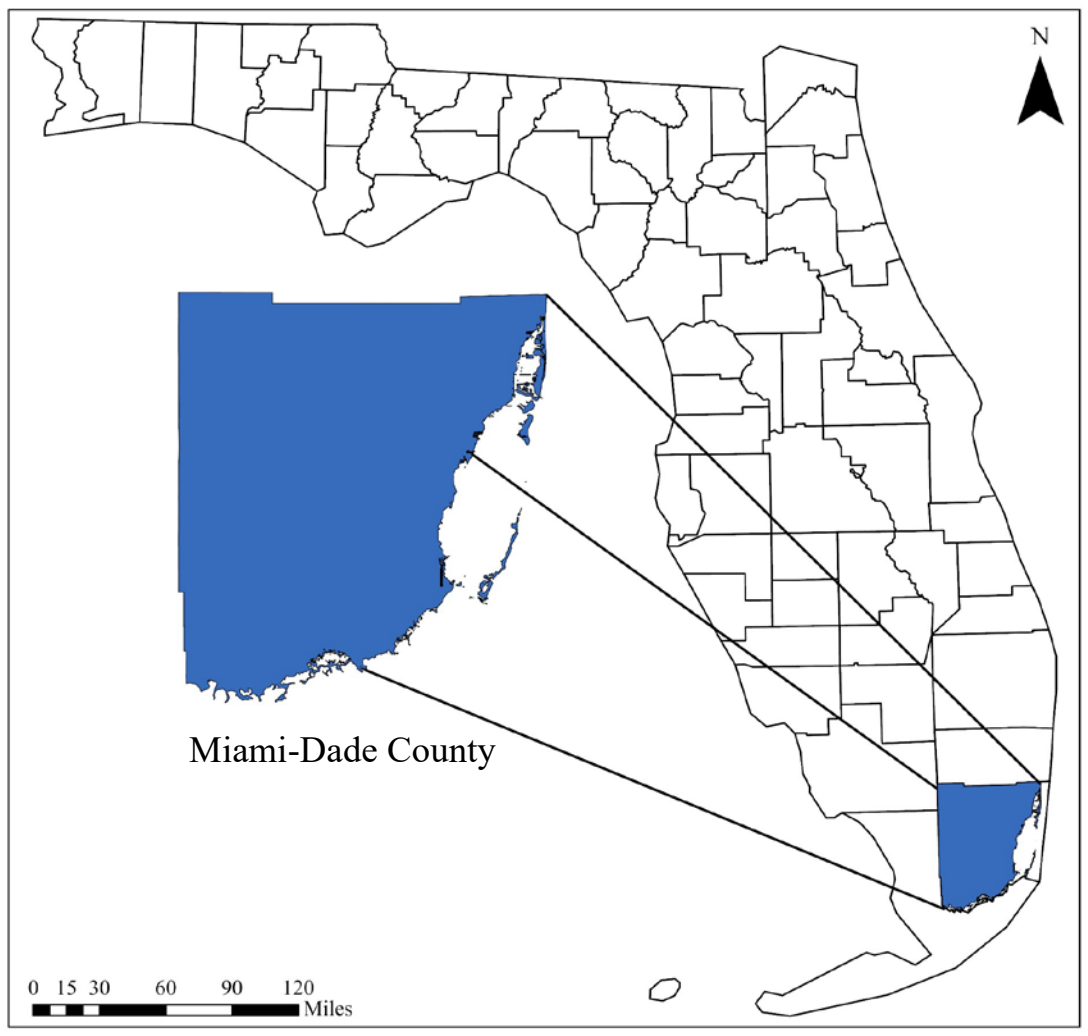

**Figure 15**. Miami-Dade County.

Using U.S. Census Bureau data (2021a), MDC covers 2,431 mi², yielding an average population density of 1,095.4 persons/mi². Assuming daily parcel demand scales with population, we estimate roughly 195 packages/mi² each day. The U.S. Postal Service (USPS), United Parcel Service (UPS), and FedEx together handle nearly 80 percent of U.S. parcel volume (Pitney Bowes, 2022). Based on their market shares, Table 7 breaks out each carrier's daily demand density. We derive carrier-specific demand growth and volatility rates from performance data in the Pitney Bowes reports (2022, 2023).

To model drone operating costs in MDC, we subdivide each square mile into multiple "delivery corridors" according to the local delivery density and fleet composition (Figure 4). Within a corridor, demand points are evenly distributed and served jointly by trucks and their onboard drones. Given daily densities of 40-70 deliveries/mi² per carrier, and assuming each truck deploys 5-15 drones, a square mile might contain 3-15 corridors. In this configuration, a drone's round-trip distance per delivery point averages 1.0-1.5 miles (from truck launch to recovery). At an operating cost of approximately $1 per delivery (Keeney, 2015), this implies drone variable costs of $0.60–$0.80 per mile.

**Table 7.** Average services metrics among USPS, UPS and FedEx.

| Logistics operators | Market share | Daily demand density | Delivery volumes per year in U.S. (Billions) | | | | Demand Growth (Daily) | Volatility rate (Daily) |
|---|---|---|---|---|---|---|---|---|
| | | | 2019 | 2020 | 2021 | 2022 | | |
| USPS | 32% | 62.4 | 5.4 | 7.3 | 6.9 | 6.7 | 0.000234 | 0.0823 |
| UPS | 25% | 48.75 | 4.3 | 4.9 | 5.3 | 5.2 | 0.000172 | 0.0826 |
| FedEx | 20% | 44.85 | 3.1 | 3.7 | 4.3 | 4.1 | 0.000304 | 0.0815 |

Source: Pitney Bowes, 2021, 2022 and 2023.

To guarantee accuracy and dependability when compared to actual logistics operations, our simulation model was thoroughly validated using case data from real-world scenarios (Pitney Bowes, 2021, 2022 and 2023). The demand growth rates and volatility of the three logistics operators can be obtained by utilizing their historical performance data. In Figure 16, GBM is used to simulate the future demand densities for the three logistics operators operating in MDC, starting with the initial densities in 2021. The horizontal axis in our simulation represents the timeline, allowing us to visualize the evolution of each logistics operator's performance over one year. The demand density lines presented in the three subfigures within Figure 16 depict the anticipated fluctuations in future demand for three logistics operators. These fluctuations are determined by considering the initial demand density, growth rate, and volatility rate, all of which are derived from historical data specific to each operator. Following that, the optimal switching points are identified by employing the proposed dynamic switching model in Section 4, as illustrated in Figure 16.

In Sections 5.2.1 to 5.2.3, it was explained that IC switching disregards switching costs, whereas dynamic deterministic and stochastic models account for them. When factoring in switching costs, some switches that IC switching deems unwarranted result in loss of benefits. It can be seen from Figure 16 that in situations involving stochastic switching, there is a potential for higher saved cost, particularly when demand growth rates are relatively high. Under such circumstances, operators may be more inclined to consider switching strategies, even if there are costs associated with each transition. Table 8 clearly demonstrates that both deterministic and stochastic models lead to greater saved cost compared to IC switching. Notably, stochastic switching offers a more realistic decision-making process by considering the uncertainty of demand density through upper and lower thresholds. Additionally, Figure 16 outlines the time periods for determining the optimal delivery service for various logistics operators, thereby illustrating the optimal delivery plan. The results of the case study indicate that our proposed dynamic switching model enhances the switching benefits for logistics operators.

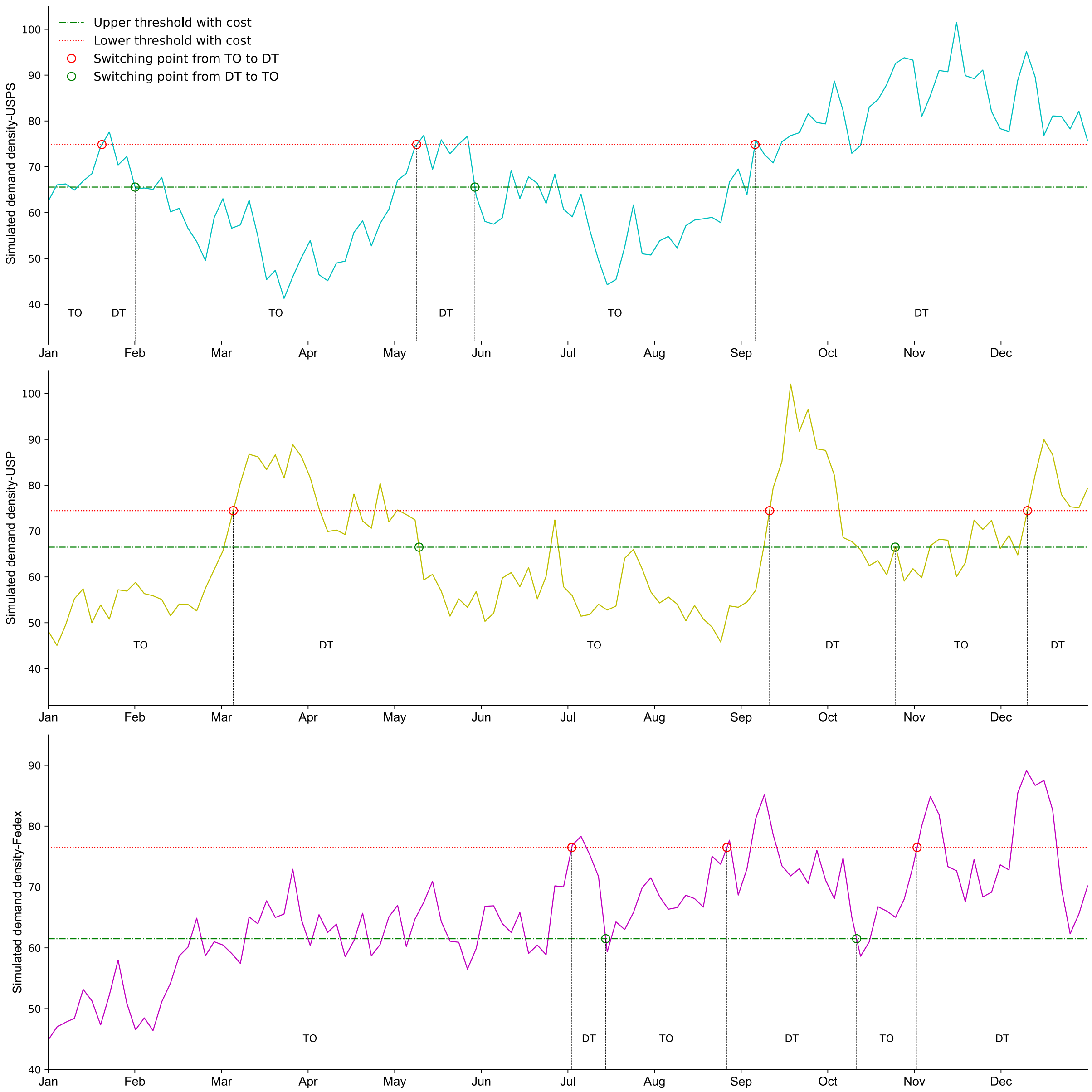

**Figure 16.** The switching timing of different logistics operators.

**Table 8.** Saved cost with switching cost of different logistics operators.

| Logistics operators | IC model ($) | Deterministic model ($) | % compared to IC model | Stochastic model ($) | $Q_L$ | $Q_H$ | % compared to IC model |
|---|---|---|---|---|---|---|---|
| USPS | 370513 | 378304 | 2.10% | 425892 | 65.1 | 74.5 | 14.95% |
| UPS | 335129 | 352712 | 5.25% | 361789 | 66.4 | 73.9 | 7.96% |
| FedEx | 130245 | 140313 | 7.73% | 168765 | 61.3 | 76.7 | 29.57% |

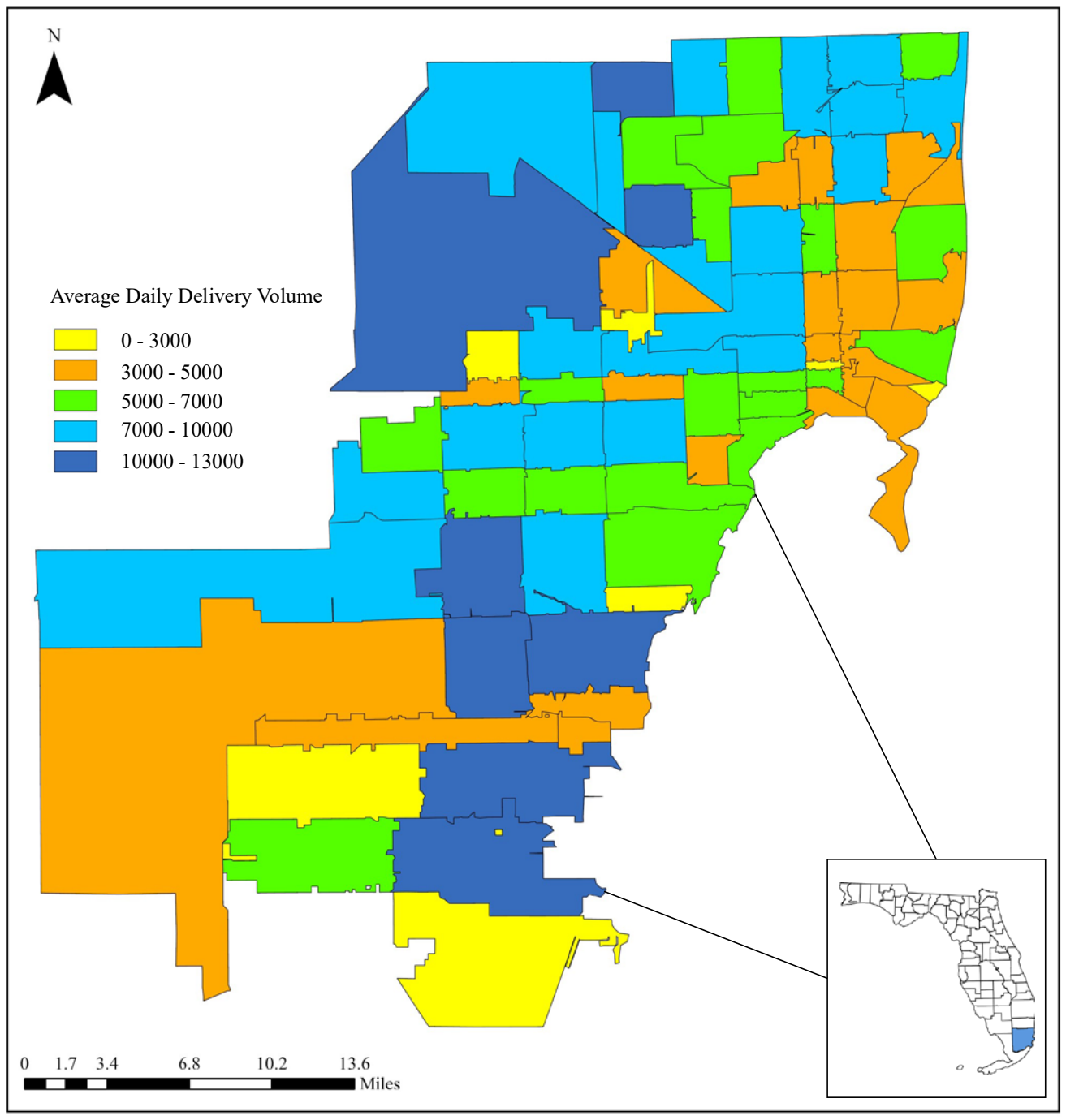

**Figure 17.** Average daily delivery volume by ZIP code tabulation area.

Based on population data sourced from the U.S. Census Bureau, 2021b, Figure 17 displays the delivery volume categorized by ZIP code tabulation areas within MDC. It can be found that the demand is extremely unbalanced within the region. To validate the benefits of our dynamic policy, we selected three days with varying demand densities based on the average demand density across the entire MDC region. Unlike the static policy, which exclusively relies on either the TO or DT fleet, the dynamic model optimizes cost savings by adapting fleet deployment based on their demand fluctuations. In the static policy shown in Figure 18(a) and 18(b), the entire region is served exclusively by the TO fleet or DT (10 drones) fleet, regardless of demand density variations. In contrast, Figure 18(c) illustrates the dynamic policy, which adapts fleet deployment to the varying demand densities within the region. Under this policy, the TO fleet is strategically deployed to areas with the lowest demand, while DT fleets with 2, 5, or 10 drones are allocated to zones based on their respective demand levels. Additionally, Table 9 presents the cost comparisons for different policies across the MDC region, demonstrating that the dynamic policy consistently results in lower costs.

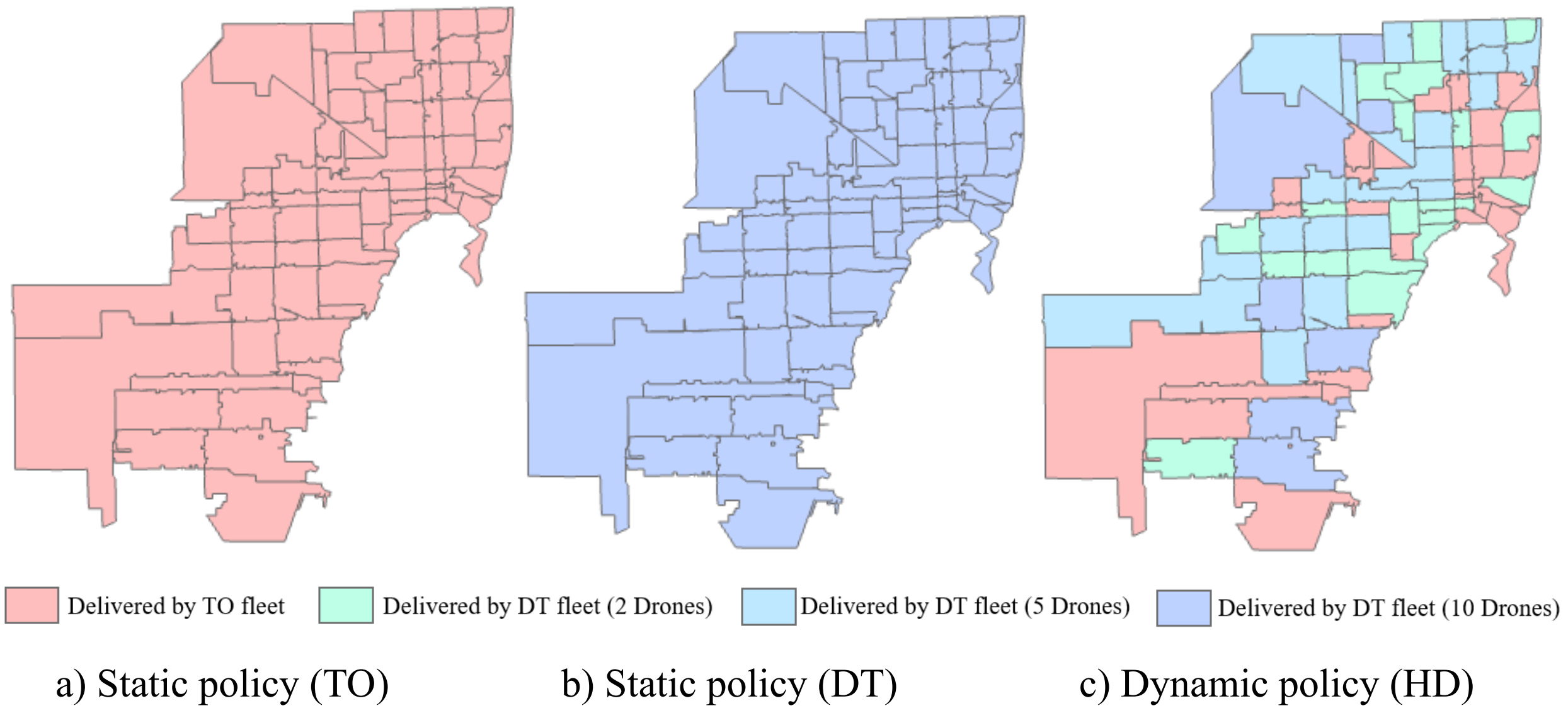

**Figure 18.** The changing of area delivered by different fleets

**Table 9:** Total cost of different policies

| | **Static policy** | | **Dynamic policy** |
|---|---|---|---|
| | **TO** | **DT** | **HD** |
| Total cost ($) | 1745824 | 1937559 | 1327940 |

## 6. Conclusions and Future Work

Incorporating drones into truck systems to address the last-mile delivery challenge has the potential to significantly improve customer access, cost efficiency, and delivery effectiveness. The adoption of such a system depends greatly on the demand for its services and the demand's variability. This paper presents a stochastic model aimed at examining the benefit for the deployment of drone-assisted truck delivery, taking into account stochastic demand densities. The study includes a comparative analysis of computational results across different switching models and a case study conducted in the Miami-Dade region involving three logistics operators. Key findings are as follows:

- A deterministic model is introduced, ensuring that the gain from each service switch surpasses the associated switching cost. With increased switching costs, the number of switching points decreases.
- Drawing from the real options approach, a stochastic model is proposed for identifying the optimal switching time. If switching cost is considered, the stochastic model results in a 31.4% reduction in costs compared to the IC model, over one year of simulated demand densities.

- The sensitivity of the total cost to the number of drones per truck is highlighted. As the number of drones increases, the cost disparity between these services grow, driven by a clustering effect that magnifies the advantages of using drones.
- A sensitivity analysis is conducted for each parameter of the stochastic switching model. Remarkably, the values and sensitivities of the switching thresholds in the stochastic switching model do not necessarily display symmetry regarding to the increasing and decreasing of various parameters.
- Furthermore, as the discount rate $\rho$, volatility rate $\sigma$, and switching cost $F^{+}$ (or $F^{-}$) increase, the entry and exit timings are delayed. In contrast, when the demand growth rate $\mu$ increases, entry occurs earlier while the exit timing is postponed.
- When operators need to dynamically allocate delivery fleets across a large region with multiple delivery zones, a multiple options switching model is proposed to dynamically transition between TO, HD, and DT. In the case of two-stage switching, the stochastic dynamic model still yields upper and lower switching thresholds for each stage.
- Our proposed model has been demonstrated in a case study for the Miami-Dade County region with data representing three major logistics operators. Through the application of our switching model, cost savings for USPS, UPS, and FedEx have significantly improved by 14.95%, 7.96%, and 29.57%, respectively. The case study includes simulations outlining the delivery service plans for each logistics operator. For different levels of demand density, we apply a dynamic policy with hybrid fleets in the MDC region. The results show that the dynamic policy reduces costs compared to the static policy.

Future enhancements could potentially address certain limitations and simplifications inherent in this study. These encompass:

- Demand density may be expressed with the mean-reverting Ornstein-Uhlenbeck process because that density fluctuates around an average value. Moreover, the demand density also has other distributions in addition to the uniform distribution, other non-uniform distributions are indicated in Appendix E.
- Switches among multiple delivery services, such as drones, e-scooters, and e-bikes, could be explored in future work to better reflect the diversity of last-mile delivery options..
- The inclusion of demand jumps could be worthwhile, since the delivery system is susceptible to abrupt demand surges or declines triggered by unforeseen events such as the Covid-19 pandemic. Incorporating demand jumps might entail raising the upper threshold and lowering the lower threshold, resulting in delayed switching.
- Although this study accounts for stochastic demand, it is worth exploring the incorporation of further stochastic parameters, such as truck speeds and weather conditions.

- The multiple options (TO, HD, and DT) can be addressed using our proposed model. To better capture multi-modal switching options, we can incorporate mothballing and activation within existing models, as detailed in Appendix F.
- Considering demand elasticity is important, as consumer demand for delivery services may vary in response to factors such as pricing, delivery speed, and service reliability. Incorporating demand elasticity into the model would enable a more responsive and adaptive optimization of delivery systems.

# Appendix A

When assuming that demand density adheres to a GBM, the evolution of demand density over time is expressed as:

$$Q(t) = Q(0)\exp\left(\left(\mu - \frac{1}{2}\sigma^2\right)t + \sigma w\right), \quad \text{(A1)}$$

Given Eq. (A1), the expectation of $Q(t)^k$ for any given value of $t$, where k is a positive real number, can be derived as follows:

$$E[Q(t)^k] = Q(0)^k e^{\left[k\mu + \frac{\sigma^2}{2}k(k-1)\right]t}. \quad \text{(A2)}$$

It is clear that all parameters $\mu$ and $\sigma$ in Eq. (A1) affect the expectation of $Q(t)^k$:

$$E_{Q(t)}\left[\int_t^{\infty} \Omega\big(Q(s)\big)e^{-\rho(s-t)}ds\right] = E_{Q(t)}\left[\int_t^{\infty}\left(\alpha_3 Q(s) + \beta_3 Q(s)^{\frac{1}{2}}\right)e^{-\rho(s-t)}ds\right]. \quad \text{(A3)}$$

in which the conditional expectation $E_{Q(t)}$ indicates that $Q(t) = Q^*$. According to Eq. (A2), by letting $k = \frac{j}{2}$, $Q(t) = Q^*$ in Eq. (A2), we obtain:

$$E_{Q(t)=Q^*}\left[Q(s)^{\frac{j}{2}}\right] = Q^{*\frac{j}{2}}e^{\omega_j s}, \quad \text{(A4)}$$

in which the coefficient $\omega_j = \frac{j}{2}\mu + \frac{\sigma^2}{2}\frac{j}{2}\left(\frac{j}{2}-1\right), j = 1,2.$

Thus, Eq. (A3) can be rewritten as:

$$E_{Q(t)=Q^*}\left[\int_t^{\infty}\left(\alpha_3 Q(s) + \beta_3 Q(s)^{\frac{1}{2}}\right)e^{-\rho(s-t)}ds\right] = \int_t^{+\infty}\left[\alpha_3 Q^* e^{(\omega_2-\rho)(s-t)} + \beta_3 Q^{*\frac{1}{2}}e^{(\omega_1-\rho)(s-t)}\right]ds, \quad \text{(A5)}$$

Eq. (A5) is integrated as:

$$\int_t^{+\infty}\left[\alpha_3 Q^* e^{(\omega_2-\rho)(s-t)} + \beta_3 Q^{*\frac{1}{2}}e^{(\omega_1-\rho)(s-t)}\right]ds = \frac{\alpha_3}{\rho-\omega_2}Q^* + \frac{\beta_3}{\rho-\omega_1}Q^{*\frac{1}{2}} = -\frac{\alpha_3}{\mu-\rho}Q^* - \frac{\beta_3}{\frac{1}{2}\left(\mu-\frac{\sigma^2}{4}\right)-\rho}Q^{*\frac{1}{2}}. \quad \text{(A6)}$$

## Appendix B

Note that $\lim_{Q\to 0+} V_0(Q) = 0$, therefore, we must have $A_0 = 0$. Similarly, since $\lim_{Q\to\infty} V_1(Q) = 0$, it follows that $B_1 = 0$ (Dixit,1994). We can therefore omit the subscripts on the remaining coefficients and express the solutions as shown in Eqs. (B1) and (B2).

$$V_0(Q) = B_0 Q^{\gamma_1}, \tag{B1}$$

and

$$V_1(Q) = A_1 Q^{\gamma_0} - \frac{\alpha_3}{\mu-\rho} Q - \frac{\beta_3}{\frac{1}{2}\left(\mu-\frac{\sigma^2}{4}\right)-\rho} Q^{\frac{1}{2}}. \tag{B2}$$

For specific choices of $\mu$, $\sigma$, and $\rho$, Eqs. (B1) and (B2) can be substituted into the value matching and smooth pasting conditions in Eq. (17)-(20) to establish four equations in four unknown variables: $B_0$, $A_1$,$Q_H$, and $Q_L$. It can be shown that the optimal solution $[B_0, A_1, Q_H, Q_L]$ is uniquely determined by solving the system of nonlinear Eqs. (B3).

$$\begin{bmatrix} A_1 {Q_L}^{\gamma_0} - B_0 {Q_L}^{\gamma_1} - \frac{\alpha_3}{\mu-\rho} Q_L - \frac{\beta_3}{\frac{1}{2}\left(\mu-\frac{\sigma^2}{4}\right)-\rho} {Q_L}^{\frac{1}{2}} + F^- \\ A_1 {Q_H}^{\gamma_0} - B_0 {Q_H}^{\gamma_1} - \frac{\alpha_3}{\mu-\rho} Q_H - \frac{\beta_3}{\frac{1}{2}\left(\mu-\frac{\sigma^2}{4}\right)-\rho} {Q_H}^{\frac{1}{2}} - F^+ \\ A_1 \gamma_0 {Q_L}^{\gamma_0-1} - B_0 \gamma_1 {Q_L}^{\gamma_1-1} - \frac{\alpha_3}{\mu-\rho} - \frac{1}{2}\frac{\beta_3}{\frac{1}{2}\left(\mu-\frac{\sigma^2}{4}\right)-\rho} {Q_L}^{-\frac{1}{2}} \\ A_1 \gamma_0 {Q_H}^{\gamma_0-1} - B_0 \gamma_1 {Q_H}^{\gamma_1-1} - \frac{\alpha_3}{\mu-\rho} - \frac{1}{2}\frac{\beta_3}{\frac{1}{2}\left(\mu-\frac{\sigma^2}{4}\right)-\rho} {Q_H}^{-\frac{1}{2}} \end{bmatrix} = 0 \tag{B3}$$

Eqs. (B3) can be solved numerically for any given set of parameter values. Let

$$K_1(\mathrm{Q}) = -\frac{\alpha_3}{\mu-\rho} Q. \tag{B4}$$

$$K_2(\mathrm{Q}) = -\frac{\beta_3}{\frac{1}{2}\left(\mu-\frac{\sigma^2}{4}\right)-\rho} Q^{\frac{1}{2}}. \tag{B5}$$

From Eqs. (B3a) and (B3c), we have

$$B_0 {Q_L}^{\gamma_1} = \frac{(1-\gamma_0)K_1(Q_L) + \left(\frac{1}{2}-\gamma_0\right)K_2(Q_L) - \gamma_0 F^-}{\gamma_1-\gamma_0}, \tag{B6}$$

$$A_1 {Q_L}^{\gamma_0} = \frac{(1-\gamma_1)K_1(Q_L) + \left(\frac{1}{2}-\gamma_1\right)K_2(Q_L) - \gamma_1 F^-}{\gamma_1-\gamma_0}. \tag{B7}$$

From Eqs. (B3b) and (B3d), we have

$$B_0 {Q_H}^{\gamma_1} = \frac{(1-\gamma_0)K_1(Q_H) + \left(\frac{1}{2}-\gamma_0\right)K_2(Q_H) + \gamma_0 F^+}{\gamma_1-\gamma_0}, \tag{B8}$$

$$A_1 {Q_H}^{\gamma_0} = \frac{(1-\gamma_1)K_1(Q_H) + \left(\frac{1}{2}-\gamma_1\right)K_2(Q_H) + \gamma_1 F^+}{\gamma_1-\gamma_0}. \tag{B9}$$

# Appendix C

This appendix discusses the expected time needed to switch from one delivery service to another. For $Q_0 = Q < Q_H$, let

$$\tau_H = \inf\{t \geq 0: Q_t = Q_H\}. \quad \text{(C1)}$$

where $\tau_H$ represents the first time that demand density $Q_t$ reaches $Q_H$. Then let $G(Q)$ denote the expected time the delivery service enters into drone-assisted truck delivery service, that can be expressed as:

$$G(Q) = E_Q[\tau_H]. \quad \text{(C2)}$$

Applying Ito's lemma (Merton, 1975), we find that $G(Q)$ satisfies the following differential equation:

$$\frac{1}{2}\sigma^2 Q^2 G(Q)'' + \mu Q G(Q)' + 1 = 0. \quad \text{(C3)}$$

This ordinary differential equation (C3) has a general solution as follows:

$$G(Q) = Q^l - \frac{1}{\eta - \frac{\sigma^2}{2}} \ln Q. \quad \text{(C4)}$$

where $l$ is a parameter which can be determined by $l = 1 - \frac{2\eta}{\sigma^2}$.

It is clear that $G(Q_H) = 0$. Assume $\eta \neq \frac{1}{2}\sigma^2$. Solving the above ODE subject to the boundary condition, the expected time needed when the drone-assisted truck delivery service is activated from truck-only delivery service can be found as:

$$G(Q) = \frac{ln\, Q_H}{\eta - \frac{\sigma^2}{2}} \left(\frac{Q}{Q_H}\right)^l - \frac{ln\, Q}{\eta - \frac{\sigma^2}{2}}. \quad \text{(C5)}$$

Hence, when the delivery service is initiated with the truck-only delivery service, the average duration for transitioning to the drone-assisted truck delivery service can be computed by obtaining the mean value of Equation (C5), as demonstrated in Eq. (23). Similarly, for $Q_0 = Q > Q_L$, let

$$\tau_L = \inf\{t \geq 0: Q_t = Q_L\}. \quad \text{(C6)}$$

where $\tau_L$ represents the first time $Q_t$ reaches $Q_L$. By a similar procedure, when the initial delivery service is drone-assisted truck delivery, the average duration for transitioning to truck-only delivery service can be computed with Eq. (24).

# Appendix D

For the special case in which the switching cost becomes zero, the upper and lower thresholds $Q_H$ and $Q_L$ converge to a single threshold $Q^*$, and then Eqs. (14-17) become:

$$V_0(Q^*) = V_1(Q^*). \tag{D1}$$

$$V_0'(Q^*) = V_1'(Q^*). \tag{D2}$$

That is,

$$\begin{bmatrix} B_0 Q^{*\gamma_1} - A_1 Q^{*\gamma_0} - \frac{\alpha_3}{\mu-\rho} Q^* - \frac{\beta_3}{\frac{1}{2}\left(\mu-\frac{\sigma^2}{4}\right)-\rho} Q^{*\frac{1}{2}} \\ B_0\gamma_1 Q^{*\gamma_1-1} - A_1\gamma_0 Q^{*\gamma_0-1} - \frac{\alpha_3}{\mu-\rho} - \frac{1}{2}\frac{\beta_3}{\frac{1}{2}\left(\mu-\frac{\sigma^2}{4}\right)-\rho} Q^{*-\frac{1}{2}} \end{bmatrix} = 0 \tag{D3}$$

Note that we must determine three unknown parameters. To find $Q^*$, we also need the following condition:

$$V_0''(Q^*) = V_1''(Q^*). \tag{D4}$$

From Eqs. (D3), we have

$$B_0 Q^{*\gamma_1} = \frac{(1-\gamma_0)K_1(Q^*) + \left(\frac{1}{2}-\gamma_0\right)K_2(Q^*)}{\gamma_1-\gamma_0}, \tag{D5}$$

$$A_1 Q^{*\gamma_0} = \frac{(1-\gamma_1)K_1(Q^*) + \left(\frac{1}{2}-\gamma_1\right)K_2(Q^*)}{\gamma_1-\gamma_0}. \tag{D6}$$

Substituting the above into (D4) leads to

$$\gamma_1(\gamma_1-1)\frac{(1-\gamma_0)K_1(Q^*) + \left(\frac{1}{2}-\gamma_0\right)K_2(Q^*)}{\gamma_1-\gamma_0} - \gamma_0(\gamma_0-1)\frac{(1-\gamma_1)K_1(Q^*) + \left(\frac{1}{2}-\gamma_1\right)K_2(Q^*)}{\gamma_1-\gamma_0} + \frac{K_2(Q^*)}{4} = 0. \tag{D7}$$

# Appendix E

We will consider one uniform distribution and two types of classical non-uniform distribution patterns:

(a) Type a: the demand density is uniform distributed in-service zone, which can be expressed as

$$Q = Q_0. \tag{E1}$$

(b) Type b: The demand density is highest at the center of service zone and decreases linearly towards the service zone boundary, which can be expressed as

$$Q = Q_0 e^{\left(-\frac{(x^2+y^2)}{(2*\omega^2)}\right)}. \tag{E2}$$

(c) Type c: The density is lowest at the center of service zone and increases linearly towards the service zone boundary, which can be expressed as

$$Q = Q_0 e^{\left(\frac{(x^2+y^2)}{(2*\omega^2)}\right)}. \tag{E3}$$

where $\omega$ is the sensitivity parameter.

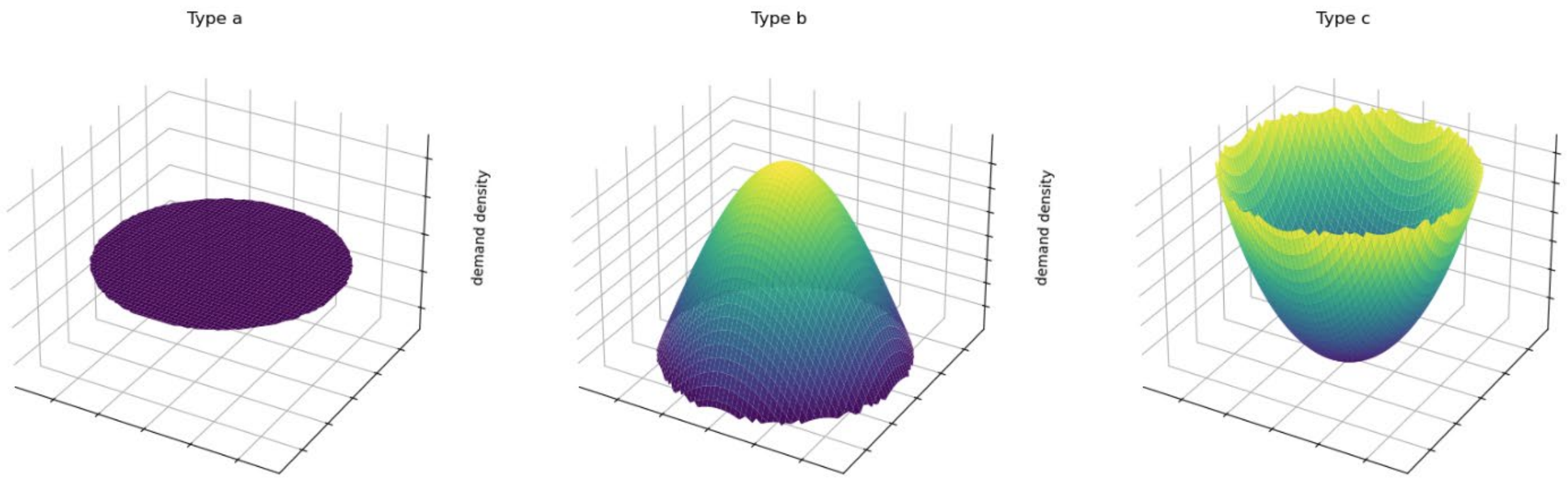

Figure E1. Three types of demand density

Figure 3 illustrates the general patterns of the three types of demand density within the service zone. For a service zone with a fixed size and layout, the average demand density must be computed, as the delivery fleet is assigned based on this metric. Consequently, the specific demand distribution has little impact on fleet assignment within the service zone. Therefore, our model adopts a uniform population density to analyze its characteristics rather than other distribution patterns. However, in future studies, we will consider different demand distributions when assigning delivery fleets to smaller zones, such as distinct delivery subzones within a single service zone.

# Appendix F

For future study, the operator has the option to entry, mothball, reactive and abandon for three options (TO, HD ans DT in multiple options switching model). Value of an idle state (TO fleets) $V_0(Q)$, active state (DT fleet) $V_1(Q)$ and mothballed state (HD fleet) $V_m(Q)$ is given by the differential equations below (Dixit,1994):

$$\frac{1}{2}\sigma^2 Q^2 V_0''(Q) + \mu Q V_0'(Q) - \rho V_0(Q) = 0, \quad \text{(F1)}$$

$$\frac{1}{2}\sigma^2 Q^2 V_1''(Q) + \mu Q V_1'(Q) - \rho V_1(Q) - \alpha_3 Q - \beta_3 Q^{\frac{1}{2}} = 0. \quad \text{(F2)}$$

$$\frac{1}{2}\sigma^2 Q^2 V_m''(Q) + \mu Q V_m'(Q) - \rho V_m(Q) - \alpha_4 Q - \beta_4 Q^{\frac{1}{2}} = 0. \quad \text{(F3)}$$

where $-\alpha_3 Q - \beta_3 Q^{\frac{1}{2}}$ is the cost difference between TO and DT and $-\alpha_4 Q - \beta_4 Q^{\frac{1}{2}}$ is the cost difference between HD and DT.

We compute the value of the firm in every state considering the available options going forward.

Equations (F4), (F5) and (F6) have a general solution of the form

$$V_0(Q) = B_0 Q^{\gamma_1}, \quad \text{(F4)}$$

$$V_1(Q) = A_1 Q^{\gamma_0} - \frac{\alpha_3}{\mu-\rho} Q - \frac{\beta_3}{\frac{1}{2}\left(\mu - \frac{\sigma^2}{4}\right) - \rho} Q^{\frac{1}{2}}. \quad \text{(F5)}$$

$$V_m(Q) = D_0 Q^{\gamma_1} + D_1 Q^{\gamma_0} - \frac{\alpha_4}{\mu-\rho} Q - \frac{\beta_4}{\frac{1}{2}\left(\mu - \frac{\sigma^2}{4}\right) - \rho} Q^{\frac{1}{2}}. \quad \text{(F6)}$$

where $B_0$, $A_1$, $D_0$, $D_1$ are constants to be determined.

At each switching point, value-matching and smooth-pasting conditions are applicable.

$$V_0(Q_H) = V_1(Q_H) - I, \quad V_0'(Q_H) = V_1'(Q_H). \quad \text{(F7)}$$

$$V_1(Q_M) = V_m(Q_M) - E_M, \; V_1'(Q_M) = V_m'(Q_M). \quad \text{(F8)}$$

$$V_m(Q_R) = V_1(Q_R) - R, \quad V_m'(Q_R) = V_1'(Q_R). \quad \text{(F9)}$$

$$V_m(Q_S) = V_0(Q_S) - E_S, \; V_m'(Q_S) = V_0'(Q_S). \quad \text{(F10)}$$

where $Q_H$, $Q_M$, $Q_R$ and $Q_S$ are the thresholds for entering from TO to DT, mothballing from DT to HD, reactiving from HD to DT and abandoning from HD to TO, respectively. $I$, $E_M$, $R$ and $E_S$ are the cost of entry, mothball, reactive and abondon.

The respective equations in (F7)-(F10) will be used to derive the values for $Q_H$, $Q_M$, $Q_R$, $Q_S$ and the coefficients $B_0$, $A_1$, $D_0$, $D_1$.

Firstly, starting with the interaction between mothballing and reactivation, at the two threshold values $Q_R$ and $Q_M$, we derive

$$
\begin{bmatrix}
-D_0 Q_R{}^{\gamma_1} + (A_1 - D_1) Q_R{}^{\gamma_0} - \frac{\alpha_3 - \alpha_4}{\mu - \rho} Q_R - \frac{\beta_3 - \beta_4}{\frac{1}{2}\left(\mu - \frac{\sigma^2}{4}\right) - \rho} Q_R{}^{\frac{1}{2}} - R \\
-D_0 \gamma_1 Q_R{}^{\gamma_1 - 1} + (A_1 - D_1) \gamma_0 Q_R{}^{\gamma_0 - 1} - \frac{\alpha_3 - \alpha_4}{\mu - \rho} - \frac{1}{2} \frac{\beta_3 - \beta_4}{\frac{1}{2}\left(\mu - \frac{\sigma^2}{4}\right) - \rho} Q_R{}^{-\frac{1}{2}} \\
-D_0 Q_M{}^{\gamma_1} + (A_1 - D_1) Q_M{}^{\gamma_0} - \frac{\alpha_3 - \alpha_4}{\mu - \rho} Q_R - \frac{\beta_3 - \beta_4}{\frac{1}{2}\left(\mu - \frac{\sigma^2}{4}\right) - \rho} Q_R{}^{\frac{1}{2}} + E_M \\
-D_0 \gamma_1 Q_M{}^{\gamma_1 - 1} + (A_1 - D_1) \gamma_0 Q_M{}^{\gamma_0 - 1} - \frac{\alpha_3 - \alpha_4}{\mu - \rho} - \frac{1}{2} \frac{\beta_3 - \beta_4}{\frac{1}{2}\left(\mu - \frac{\sigma^2}{4}\right) - \rho} Q_M{}^{-\frac{1}{2}}
\end{bmatrix} = 0, \qquad \text{(F11)}
$$

Secondly, we consider the value-matching and smooth-pasting conditions for new investment:

$$
\begin{bmatrix}
-B_0 Q_H{}^{\gamma_1} + A_1 Q_H{}^{\gamma_0} - \frac{\alpha_3}{\mu - \rho} Q_H - \frac{\beta_3}{\frac{1}{2}\left(\mu - \frac{\sigma^2}{4}\right) - \rho} Q_H{}^{\frac{1}{2}} - I \\
-B_0 \gamma_1 Q_H{}^{\gamma_1 - 1} + A_1 \gamma_0 Q_H{}^{\gamma_0 - 1} - \frac{\alpha_3}{\mu - \rho} - \frac{1}{2} \frac{\beta_3}{\frac{1}{2}\left(\mu - \frac{\sigma^2}{4}\right) - \rho} Q_H{}^{-\frac{1}{2}}
\end{bmatrix} = 0, \qquad \text{(F12)}
$$

These conditions at the scrapping threshold $Q_S$ become

$$
\begin{bmatrix}
(D_1 - B_1) Q_S{}^{\gamma_1} + D_1 Q_S{}^{\gamma_0} - \frac{\alpha_4}{\mu - \rho} Q_S - \frac{\beta_4}{\frac{1}{2}\left(\mu - \frac{\sigma^2}{4}\right) - \rho} Q_S{}^{\frac{1}{2}} + E_S \\
(D_1 - B_1) \gamma_1 Q_S{}^{\gamma_1 - 1} + D_1 \gamma_0 Q_S{}^{\gamma_0 - 1} - \frac{\alpha_4}{\mu - \rho} - \frac{1}{2} \frac{\beta_4}{\frac{1}{2}\left(\mu - \frac{\sigma^2}{4}\right) - \rho} Q_S{}^{-\frac{1}{2}}
\end{bmatrix} = 0, \qquad \text{(F13)}
$$

Tha above equations can be solved numerically for any given set of parameter values.

# Appendix G

The expected cost for truck-only delivery, as originally derived by Campbell et al. (2017), is briefly summarized here. Each truck's delivery zone is assumed to be compact and approximately covered by a swath of width w. The truck follows an L1 metric as it travels along the swath. On average, the distance between adjacent stops horizontally is $\frac{w}{3}$, and longitudinally it is $\frac{1}{Qw}$. The expected last-mile distance per delivery point $D_{TO}$ is the sum of both, shown in Eq. (G1)

$$D_{TO} = \frac{w}{3} + \frac{1}{Qw}, \tag{G1}$$

The optimal swath width $w^*$ that minimizes the expected distance of TO is given by the first derivative of the expected distance per delivery point $D_{TO}$ with respect to the swath width $w_{to}^*$.

$$w^* = \sqrt{\frac{3}{Q}}, \tag{G2}$$

From the depot, the truck travels a linehaul distance to the first delivery point, which is distributed randomly and independently in the delivery area. In Figure 3, the truck travels the same linehaul distance to return to the depot after serving all its delivery points. In practice, a truck does not take a direct route from the depot to the delivery area. Therefore, to more precisely estimate the linehaul distance, a circuity factor is factored in. The expected straight linehaul distance $L$ in a service zone $A$ can be approximated by Eq. (G3).

$$L = 2 * \phi * \nu * \sqrt{\frac{A}{\pi}}, \tag{G3}$$

where $\phi$ is circuity factor, $\nu$ is the adjust factor for line-haul distance based on the radius of service zone.

The number of delivery points per route $m_{TO}$ depends on the given route time $T$, the truck travel time of line-haul $\frac{L}{V_l}$ and last mile $\frac{D_{TO}}{V_t}$, truck stop time:

$$m_{TO} = \frac{\left(T - \frac{L}{V_l}\right)}{\left(\frac{D_{TO}}{V_t} + d_t\right)}, \tag{G4}$$

where $V_l$ is the truck linehaul speed, $V_t$ is the truck tour speed in delivery area, and $d_t$ is average stopping time at a delivery point.

The total travel distance for TO includes the linehaul distance between depot and delivery area and the local distance to each customer in the delivery area. The optimal cost of TO per delivery point $E_{TO}$ is the sum of truck variable cost and fixed truck stop cost:

$$E_{TO} = C_t^o * \left(D_{TO} + \frac{L}{m_{TO}}\right) + S_t, \tag{G5}$$

where $C_t^o$ is the variable cost of a truck per unit distance and $S_t$ is the fixed truck stop cost per delivery point.

# Appendix H

To derive the optimal swath width $w^*$ that minimizes the expected drone travel distance per delivery point, we begin with the approximate expression provided in Eq. (19):

$$D_{DT}^{d} = \frac{2n}{n+1}\sqrt{\left(\frac{w}{3}\right)^2 + \left(\frac{n+1}{2Qw}\right)^2}, \tag{H1}$$

Since the multiplicative factor $\frac{2n}{n+1}$ does not affect the optimal value of $w$, we define the objective function to minimize as:

$$\frac{\partial D_{DT}^{d}}{\partial w} = \sqrt{\frac{w^2}{9} + \frac{(n+1)^2}{4Q^2w^2}}, \tag{H2}$$

Let $g(w) = \frac{w^2}{9} + \frac{(n+1)^2}{4Q^2w^2}$. Taking the derivative and setting it to zero:

$$\frac{\partial g(w)}{\partial w} = \frac{2w}{9} - \frac{2(n+1)^2}{4Q^2w^3}, \tag{H3}$$

Solving for $w$ gives the optimal swath width:

$$w^* = \sqrt{\frac{3(n+1)}{2Q}}, \tag{H4}$$